\documentclass[letterpaper,twocolumn,10pt]{article}
\usepackage{usenix}

\usepackage{tikz}
\usepackage{amsmath}

\usepackage{filecontents}

\usepackage{tikz}
\usepackage{amsmath}

\usepackage{filecontents}

\usepackage[noend]{algpseudocode}
\usepackage{float}
\usepackage{tabularx}
\usepackage{booktabs}
\usepackage[linesnumbered,ruled,vlined]{algorithm2e}
\usepackage{multirow}
\usepackage{booktabs}
\usepackage{pifont}
\usepackage{soul}
\usepackage[inline]{enumitem}
\usepackage{colortbl}
\usepackage{comment}
\usepackage{amsthm}
\usepackage{amssymb}
\usepackage{adjustbox}
\usepackage{xurl}

\usepackage{tikz}
\usepackage{tikz-qtree}
\usepackage{forest}
\usepackage{listings}
\usepackage[most]{tcolorbox}
\usepackage{xspace}

\newtcolorbox{promptbox}[1][]{%
  enhanced jigsaw,
  breakable, %
  colback=gray!5,
  colframe=black,
  width=\textwidth,
  parbox=false,
  halign=flush left,
  fonttitle=\bfseries,
  title=Prompt,
  fontupper=\small,
  #1
}

\usepackage{circledsteps}
\tikzset{/csteps/inner xsep=4pt, /csteps/inner ysep=4pt}
\newcommand{\cnum}[1]{\CircledText{\scriptsize #1}}

\newcommand{\fig}[1]{Fig.~\ref{#1}}

\newcommand{\tab}[1]{Tab.~\ref{#1}}
\newcommand{\alg}[1]{Alg.~\ref{#1}}

\newcommand{\app}[1]{App.~\ref{#1}}
\newcommand{\sect}[1]{§\ref{#1}}
\newcommand{\sys}{MiniScope\xspace}

\begin{document}

\date{}

\title{\sys: Authorizing Agents with Least-Privilege Permissions}

\author{
{\rm Jinhao Zhu \enspace Xiao Huang \enspace Kevin Tseng \enspace
Gil Vernik$^\dagger$ \enspace Shishir Patil \enspace Vivian Fang \enspace Raluca Ada Popa}\\[0.5em]
UC Berkeley \qquad $^\dagger$IBM Research
} %

\maketitle

\begin{abstract}

AI agents are increasingly granted autonomous access to sensitive user data and third-party services, making effective permission management a critical security challenge.
Existing permission models, however, typically rely on flat permission structures that fail to balance security with usability: fine-grained confirmation induces user fatigue, while coarse-grained or persistent approval leads to overprivileged agents. 
To address this tradeoff, we propose a task-centric, hierarchical permission model that treats an agent as a delegate operating within a task-specific role instead of requiring a separate permission decision for every tool call.
Building on this model, we present \sys, an end-to-end permission system for agents that automates permission-hierarchy discovery and enforces contextual least privilege at runtime.
Our evaluation shows that \sys reduces simulated permission confirmations by 43.4\%--89.4\% for cautious and typical personas relative to per-tool prompting and mitigates all privilege-escalation attacks with negligible impact on utility and runtime.
Applied to real-world deployments, \sys further uncovers six overprivileged connector configurations in ChatGPT and Claude.

\end{abstract}

\section{Introduction}

Large language models (LLMs) are evolving from conversational assistants into delegated agents that act with real authority over users' accounts and devices.
ChatGPT~\cite{ChatGPT}, Claude~\cite{Claude}, and Gemini~\cite{Gemini} already integrate with external services through connectors~\cite{ChatGPT-conn, Claude-conn, Gemini-workspace}, giving agents access to capabilities such as reading email, managing calendars, searching cloud storage, and updating workspace data.
A newer wave of autonomous agents~\cite{openclaw, chagpt-operator} extends this delegation further, operating across browsers and the user's entire computer~\cite{chagpt-cua}.
Users are no longer only sharing information with LLMs; they are authorizing agents to exercise sensitive capabilities on their behalf.

However, authorizing an agent with broad capabilities creates real risk.
Agents can diverge from user intent because of indirect prompt injection~\cite{greshake2023not, liu2023prompt}, model unreliability~\cite{rawte2023survey, zhang2025siren}, or ambiguity in the user's request~\cite{yi2025privacy}.
Recent incidents make this concrete: an OpenClaw agent deleted a user's entire email inbox despite being told to confirm before acting~\cite{meta-email-delete}, while another AI agent reportedly deleted an entire company's database~\cite{pocketos-delete}.

The core problem is not only that agents can make mistakes, but that today's agents are often granted permissions far broader than what a task actually requires \cite{ms-mcp-risk, owasp-excessive}.
Ideally, an agent should be authorized with exactly the capabilities required for its current task, and nothing more.
Consider Alice, who every morning asks her assistant: \textit{``Summarize my unread emails and add calendar events for anything urgent.''}
Fulfilling this task requires only listing and reading messages in her inbox, and reading and creating events on her calendar.
Under such scoping, the incidents above would be structurally impossible: deleting the inbox or the database would lie outside the permissions granted for the task.

Agentic systems therefore need a permission model that enforces least privilege for each task.
Existing approaches, however, largely adapt permission models designed for conventional mobile applications instead of rethinking authorization for general-purpose agents.
\textit{Install-time approval}, used by most connector-based systems~\cite{ChatGPT-conn, Claude-conn}, asks users to approve a set of service permissions when connecting the service and leaves those permissions available to the agent across subsequent tasks.
This resembles the install-time permission model used by early Android systems~\cite{felt2011android}.
\textit{Per-call confirmation}~\cite{Claude, anthropic-mcp} moves the decision closer to capability use by asking the user to approve each tool invocation before it executes.
This shift parallels the evolution of Android toward runtime permissions in Android 6.0 and one-time permissions in Android 11~\cite{android-access-control}.
Neither pattern, however, balances security with usability.
Install-time or persistent approval minimizes ongoing friction but leaves permissions active across tasks that do not require them, while per-call confirmation avoids such ambient authority at the cost of repeated interruptions and confirmation fatigue~\cite{kirwan2020repetition}.
The persistence of this tradeoff reflects a deeper problem: both patterns retain an app-centric permission model, treating the agent as a persistent principal and its capabilities as a flat list of disconnected tools.
Such a model does not capture that a general-purpose agent assumes different roles, and therefore requires different authority, from one task to another.
Resolving this mismatch requires rethinking not only when permissions are requested, but how the granted authority is scoped.

Our key insight is that an agent should be authorized like a human delegate: through a role that captures the permissions needed to complete the user's task, rather than through a flat set of individual tool permissions.
Like organizational roles, agent roles form a hierarchy, with broader roles for composite tasks and narrower roles for subtasks.
We therefore depart from flat, per-tool authorization and propose a
\textbf{task-centric, hierarchical permission model} for agents.
Building on this model, we present \textbf{\sys}, a least-privilege permission system that adapts classical \textit{hierarchical role-based access control}
(RBAC)~\cite{sandhu1998role} to the agentic setting.
The adaptation is \textit{contextual}~\cite{tsai2025contextual}: whereas classical RBAC binds a principal to a role for an extended period, \sys activates only the roles selected for the current task, even when the user has authorized broader roles for reuse.
This delivers least privilege at the granularity of permission roles, and can be complemented with finer-grained runtime enforcement (\sect{sec:discussion}).

To realize \sys, we first develop \textbf{an automated pipeline for permission hierarchy construction} with two pathways.
The first reconstructs a hierarchy from service scopes.
OAuth and similar authorization frameworks manage service capabilities through scopes, which bundle related operations together (for example, a \texttt{gmail.readonly} scope permits listing and reading messages).
Scopes encode a co-occurrence signal about which capabilities belong together, but the hierarchical structure relating scopes, operations, and task-level roles is not explicit; it must be reconstructed.
The second pathway learns a hierarchy from historical traces, supporting cases when users want a permission structure tailored to their own workflows, or when the service exposes too little scope structure for reconstruction.
Inspired by role mining~\cite{li2007role} in the RBAC literature, \sys mines the hierarchy from past task executions, discovering latent task-level roles that cluster tools by how they are actually used.

\begin{figure}[!t]
    \centering
    \includegraphics[width=1\linewidth]{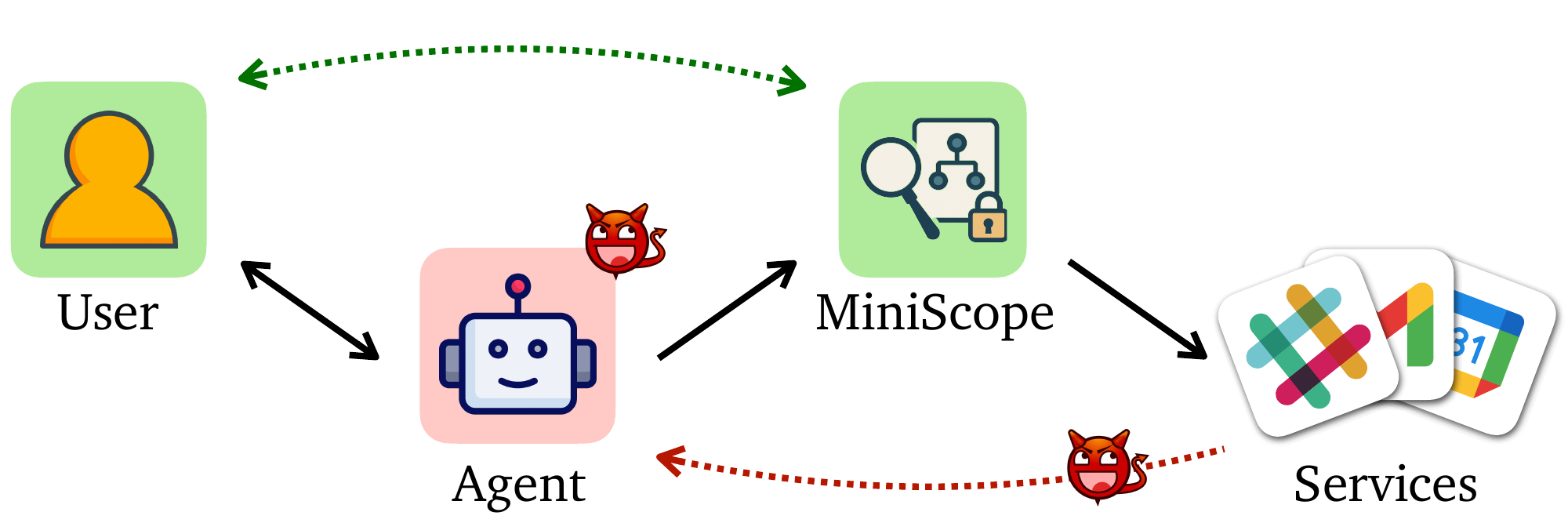}
    \caption{\sys Architecture. Red indicates untrusted. Green indicates trusted.}
    \label{fig:arch}
\end{figure}

With this hierarchy, we build \textbf{an end-to-end framework that authorizes tool-calling agents with least privilege} at runtime (\fig{fig:arch}).
The framework maps the execution plan derived from each user request to a least-privilege set of roles and enforces the resulting permission boundary throughout the task.
Specifically, it maintains two layers of permission state: \textit{contextual least-privilege permissions} contain the roles selected for the current task, while \textit{user-authorized permissions} record the roles the user has approved for possible reuse.
For each task, \sys activates only the contextual roles and prompts the user, using familiar authorization choices inspired by traditional mobile platforms~\cite{apple-access-control, android-access-control}, for any selected role not covered by the user-authorized permissions.
This separation lets \sys deliver contextual least privilege without overwhelming the user: user-authorized permissions reduce repeated prompts across tasks, while contextual permissions ensure that only the roles required for the current task are active.

\textbf{\textit{Evaluation summary.}}
We build an evaluation environment with the tool interfaces and OAuth permission hierarchies of 10 real-world applications, capturing the complexity of production deployments.
We first compare \sys with an LLM-assisted permission inference baseline and find that it produces overprivileged permissions in 1.7\%--34.7\% of cases, with the rate growing as tasks become more complex.
In a user-trace simulation across Gmail, Google Calendar, and Google Drive, \sys reduces confirmation prompts by 43--89\% relative to per-tool prompting for security-conscious and typical personas.
On the Workspace benchmark adapted from AgentDojo~\cite{debenedetti2024agentdojo} and AgentDyn~\cite{li2026agentdyn}, \sys reduces the attack success rate to 0.8\% overall and 0\% for privilege-escalation attacks (\sect{sec:sec-game}), adding only 0.14 seconds of average permission-module overhead per task.
Finally, applying \sys's least-privilege solver to deployed ChatGPT and Claude connectors uncovers \textit{six overprivileged configurations} (\sect{sec:overprivilege}), concrete evidence that hierarchical permissions surface real security gaps in production systems.

\section{System Overview}
\label{sec:overview}

\sys focuses on the user-agent-service model, a paradigm widely adopted in real-world agent deployments~\cite{apple-intelligense, pixel-ai, ChatGPT-conn, Claude-conn}.
As shown in \fig{fig:arch}, this model consists of three parties: the user, the LLM agent, and the services (applications).
In a typical workflow, the agent translates the user's natural-language request into tool calls to connected services, processes their responses, and returns the result.
Interactions may involve multiple rounds of communication between the user and the agent or between the agent and the services to gather information, execute data-dependent operations, or refine results based on user feedback.
Modern agent systems may also include additional components such as memory, skills, terminal access, or browser/computer-use capabilities; we focus on the user-agent-service model because it captures the core permission-delegation problem underlying these richer deployments.
In this paper, we study permission delegation for a single agent acting on behalf of one user, and leave multi-agent settings, including delegation between agents, to future work.
These scenarios are briefly discussed in \sect{sec:discussion}.

\subsection{Threat Model}
\label{sec:threat-model}

Our security goal is to ensure that the agent can execute only tool calls authorized by the permissions delegated to it for the current task or session.
The delegated permissions are derived from the user's task and the service's permission hierarchy, and are granted only after user approval.
Once permissions are granted, \sys enforces them as a runtime capability boundary: if the agent attempts to invoke a tool outside the granted permissions, the call must be rejected before reaching the service.
This goal is independent of why the agent issues the tool call; the request may arise from normal planning, model hallucination, ambiguous user instructions, or adversarially injected data.
With this goal and system model, we make the following assumptions for each entity in the system.

\smallskip\noindent\textbf{User.}
We trust the user's high-level intent and treat the user as the authority for permission delegation, as in many real-world deployments~\cite{anthropic-mcp, magentic-ui, mozannar2025magentic, shavit2023practices, wu2025isolategpt}.
When \sys requests additional permissions, the user's decision determines whether those permissions may be delegated to the agent.
However, we do not assume that the full prompt is trusted: a user may include untrusted content from documents, web pages, or other external sources.
Such content may be benign, ambiguous, or adversarial, and may cause the agent to request or invoke capabilities beyond the user's intended task~\cite{chen2024agentpoison,xu2026storage, greshake2023not}.

\smallskip\noindent\textbf{Model.}
We make no trust assumptions about the underlying LLM used by the agent, which may be a proprietary model from a major provider~\cite{Gemini, ChatGPT, Claude} or a custom model hosted by the user.
The model may hallucinate, misinterpret the user's task, be influenced by untrusted context, or issue tool calls that exceed the permissions needed for the task~\cite{liu2023prompt, shi2025prompt}.
Accordingly, \sys does not rely on the model to report what permissions it needs or to remain within a previously approved execution plan.

\smallskip\noindent\textbf{Services.}
For the services connected to the agent, \sys assumes that their tool call implementations follow the intended specifications.
For example, if a tool is designed to read an email, its implementation should actually perform a read operation on the target email rather than delete it.
This assumption does not require every service to implement all tools correctly.
Rather, our security guarantee applies only to services, or specific service methods, whose behavior matches their stated specification.
When \sys constructs a permission hierarchy from a service-provided mapping, such as OAuth scopes, we additionally assume that the service provider's permission specification correctly defines which tools each permission authorizes.
Alternatively, when the hierarchy is learned from historical tool-use traces, \sys does not require a service-provided permission mapping.
Finally, tool call responses from services are treated as untrusted, since they may contain external or attacker-controlled content.

\smallskip\noindent\textbf{Non-goals.}
\sys enforces permission boundaries at the granularity of the permission hierarchy.
It prevents the agent from expanding its capabilities beyond the permissions granted, but it does not verify the semantic correctness of tool calls that are already within those permissions.
For example, if the agent is granted permission to send email, \sys does not determine whether the agent's choices of recipient, subject, and body reflect the user's intent.
Such within-permission misuse may arise from prompt injection, hallucination, ambiguity, or malicious model behavior, but it is outside \sys's enforcement boundary.
Accordingly, \sys is not a general defense against prompt injection, jailbreaking, or attacks that target the model itself~\cite{wei2023jailbroken,zou2023universal}. 
We discuss possible extensions for finer-grained protection in~\sect{sec:fine-grained}.
We also provide a detailed discussion of in-scope and out-of-scope attacks in \sect{sec:analysis}.
Finally, \sys does not aim to defend against denial-of-service attacks~\cite{lau2000distributed} initiated by either the model provider or the service provider, since these parties are typically incentivized to maintain service availability.

\begin{figure}[!t]
    \centering
    \adjustbox{margin=0.5em,width=\linewidth,frame,center}{
    \fontsize{10pt}{12pt}\selectfont
    \begin{minipage}{\linewidth}
        \begin{enumerate}
            \item The challenger selects a set of services $\mathcal{S}$ and a permission hierarchy $\mathcal{H}$ over the tool calls exposed by these services.
            \item The adversary chooses the model $f_{\mathrm{LLM}}$ and valid service states $\mu = (\mu_1,\ldots,\mu_n)$ for the services in $\mathcal{S}$.
            \item The agent is initialized as
    \[
        A = \mathsf{Agent}\big(f_{\mathrm{LLM}}, \mathcal{S}(\mu)\big).
    \]
            \item For each round $t$, the challenger issues a user task, and \sys maintains a delegated permission state $\rho_t$ 
            \item The challenger interacts with the agent $A$ through \sys for $N$ rounds, producing an execution trace $\tau$ of tool calls executed on services.
            \item The adversary wins if any tool call $c \in \tau$ is executed at time $t$ while $c \notin \mathsf{Auth}_{\mathcal{H}}(\rho_t)$.
        \end{enumerate}
    \end{minipage}
    }   
    \caption{Security game for \sys}
    \label{fig:game}
\end{figure}

\subsection{Security Game}
\label{sec:sec-game}

We formalize \sys's security goal as preventing privilege escalation within a hierarchical permission model.
In such a model, permissions are ordered by the capabilities they authorize: higher-level permissions grant broader access, while lower-level permissions grant narrower access.
A vertical privilege escalation~\cite{owasp-privilege-escalation} occurs when an agent authorized under a narrower permission boundary successfully executes a tool call that requires a broader permission not delegated by the user.
For example, an agent granted read-only email access escalates privilege if it successfully invokes deletion.
\sys focuses on preventing capability expansion across permission boundaries, but not misuse of already authorized capabilities. 
For example, an agent may send an email to the wrong recipient while the send-email permission is active.

To define our security game, we first introduce the following notation.
Let $\mathcal{S} = \{S_1,\ldots,S_n\}$ denote the set of services connected to the agent.
Let $\mathcal{H}$ denote the permission hierarchy over the tool calls exposed by these services.
Depending on how the hierarchy is constructed, $\mathcal{H}$ may decompose into per-service hierarchies or capture permission relationships across services.
The hierarchy defines which tool calls are authorized by each permission.
Let $\mu = (\mu_1,\ldots,\mu_n)$ denote the service states, where $\mu_i$ is a valid state of service $S_i$.
These states may contain untrusted or adversarial content that can later be returned to the agent by tool calls.
Let $f_{\mathrm{LLM}}$ denote the underlying language model, and let
\[
    A = \mathsf{Agent}\big(f_{\mathrm{LLM}}, \mathcal{S}(\mu)\big)
\]
denote the resulting tool-calling agent.
During an interaction, \sys maintains a delegated permission state $\rho_t$ at each time $t$.
This state records the permissions currently delegated to the agent.
Given $\rho_t$ and the permission hierarchy $\mathcal{H}$, let $\mathsf{Auth}_{\mathcal{H}}(\rho_t)$ denote the set of tool calls authorized at time $t$.
Finally, let $\tau$ denote the execution trace, consisting only of tool calls that are actually executed on services.

We define our security game in \fig{fig:game}.
In this game, the challenger plays the role of the user and selects the connected services $\mathcal{S}$ together with the permission hierarchy $\mathcal{H}$.
The adversary chooses the underlying model $f_{\mathrm{LLM}}$ and the valid service states $\mu$.
This captures both an untrusted model and untrusted service content that may influence the agent's future tool requests.
The challenger then interacts with the agent through \sys for $N$ rounds, during which \sys updates $\rho_t$ according to the user's permission decisions and records in $\tau$ the tool calls actually executed on services.
The adversary wins if any tool call $c \in \tau$ is executed at time $t$ while $c \notin \mathsf{Auth}_{\mathcal{H}}(\rho_t)$.
We say the system is secure against permission-level privilege escalation if the adversary wins with only negligible probability.

This guarantee is stated with respect to a fixed permission hierarchy $\mathcal{H}$.
Importantly, errors in one part of the hierarchy are localized: if a permission relationship is incorrect, the resulting least-privilege selection may be affected for tasks that rely on that part of the hierarchy, but enforcement remains unchanged for tasks whose required permissions lie in unaffected parts of the hierarchy.

\section{Finding Permission Hierarchies}

\label{sec:hierarchy}

This section describes how \sys constructs permission hierarchies for agent authorization.
We first ground the choice of hierarchy in two structural observations about how agents use tools (\sect{sec:tools-to-roles}).
We then present two construction pathways: reconstruction from a service's OAuth scopes (\sect{sec:hierarchy-oauth}), and mining from a user's prior interaction traces (\sect{sec:hierarchy-traces}).
The two pathways are complementary: scope reconstruction applies whenever a service exposes sufficient scope structure, while trace mining handles services that do not, or users who want a hierarchy tailored to their own workflows.

\subsection{From Flat Tools to Hierarchical Roles}
\label{sec:tools-to-roles}

Existing agentic permission systems treat an agent's capabilities as a flat list of individual tools. 
When combined with per-call confirmation, this model requires a permission decision for every tool invocation, creating fatigue that drives users toward permissive defaults~\cite{kirwan2020repetition, cc-auto-mode}.
The flat view, however, does not match how agents actually use tools.

We make two observations about agent tool use that motivate a hierarchical structure.
\textit{Intra-task co-occurrence:} the tools an agent invokes within a single task cluster into recurring combinations dictated by the task itself.
Reading unread email entails listing the inbox and then reading each message; adding a calendar event entails reading existing events and then creating a new one.
These groupings are not arbitrary. 
They reflect the structural decomposition of the task into the operations needed to complete it.
\textit{Inter-task consistency:} users make stable permission decisions across recurring instances of the same task.
Alice's email triage recurs daily, and she is willing to grant the same capabilities each time.
Together, these observations identify the \textit{task-level role}, rather than the individual tool, as the natural unit of permission: a bounded set of capabilities required to perform a coherent task on the user's behalf.
Because tasks decompose into subtasks, these roles naturally form a hierarchy: narrower sub-roles capture smaller subtasks, while broader roles capture composite tasks.

Our key idea is to treat an agent as a human delegate acting on the user's behalf and authorize it through a role that captures the capabilities required for its assigned task.
Following this idea, we adapt classical \textit{hierarchical role-based access control} (RBAC)~\cite{sandhu1998role} to the agentic setting.
In \sys, each role bounds the capabilities needed for a task or subtask, and the hierarchy orders roles from narrower subtasks to broader composite tasks.
Unlike traditional RBAC, where roles are assigned to a principal for long periods, \sys instead assigns an agent's role per task and changes it with context, allowing the same agent to act as an email triager in one request and a calendar manager in the next.
This design reflects prior findings that agentic permissions are inherently contextual~\cite{tsai2025contextual, wu2025towards}.
We next show how \sys derives these hierarchies in practice.

\subsection{Hierarchies from OAuth Scopes}
\label{sec:hierarchy-oauth}

Constructing a permission hierarchy requires a source of structure: some signal about which tools should be grouped together and which groups are broader than others.
One existing source of such structure is OAuth.

OAuth~\cite{rfc6749} is an open standard authorization protocol that allows third-party applications to access protected resources on behalf of users without exposing their credentials.
Within OAuth, scopes define the permissions granted to an application: each scope authorizes a set of API methods under a single grant.
Although services usually expose scopes as a flat list, the method coverage of those scopes reveals an implicit hierarchy: a scope that authorizes a strict superset of another scope's methods represents broader authority.
\sys constructs the hierarchy from this method coverage relation.
For each scope $s_i$, let $O_i$ denote the set of API methods authorized by $s_i$.
For any two scopes $s_1$ and $s_2$, if $O_1 \subset O_2$, then $s_2$ grants broader authority than $s_1$, and \sys places $s_1$ as a child of $s_2$ in the hierarchy.
The resulting graph orders scopes by increasing authority from children to parents.
The full reconstruction algorithm appears in \alg{alg:build-scope-tree}.

\begin{figure}[!t]
    \centering
     \includegraphics[width=1\linewidth]{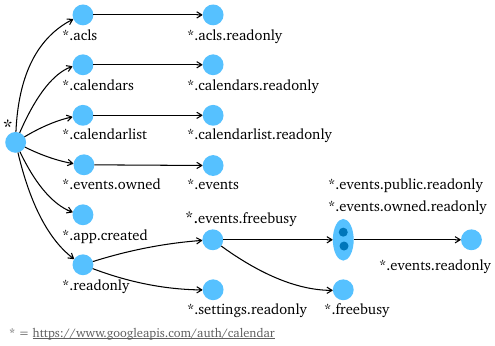}
    \caption{Google Calendar's Permission Hierarchy}
    \label{fig:gcal_tree}
\end{figure}

\fig{fig:gcal_tree} shows the hierarchy reconstructed for Google Calendar, with each node labeled by Google's original scope name.
The example shows that scope names can suggest a different hierarchy from the one implied by method coverage: \path{*.events.readonly} appears broader than \path{*.events.public.readonly} and \path{*.events.owned.readonly}, but the reconstructed hierarchy reverses this relation because the latter scopes authorize more API methods.

\subsection{Hierarchies from User Traces}
\label{sec:hierarchy-traces}

When service-defined permission groups are unavailable, or when users prefer a hierarchy tailored to their workflows, \sys derives one from historical user traces.
These traces provide a bottom-up signal: tools that repeatedly co-occur across similar tasks often form reusable permission groups.
For example, email triage may repeatedly combine listing messages with reading selected messages, while scheduling may combine checking availability with creating calendar events.
We therefore adapt role mining~\cite{li2007role}, originally developed in the RBAC literature, to reconstruct permission hierarchies from observed user-agent workflows.

\textbf{Classical role mining} takes as input a user-permission relation and seeks to recover a set of roles that compactly explains this relation.
Formally, the input is an access-control configuration $\Gamma = \langle U, P, UP\rangle$, where $U$ is a set of users, $P$ is a set of permissions, and $UP \subseteq U \times P$ records which users have which permissions.
The output is a role-based configuration $\Gamma' = \langle U, P, R, UA, PA\rangle$, where $R$ is a set of roles, $UA \subseteq U \times R$ assigns users to roles, and $PA \subseteq R \times P$ assigns permissions to roles.
The goal is for the permissions induced by $UA$ and $PA$ to reproduce or approximate the original relation $UP$ using a compact set of roles.
In hierarchical role mining, the output additionally includes a role hierarchy $RH$, where broader roles include the permissions of narrower roles.

\textbf{Agentic adaptation.} 
Classical role mining assumes persistent users with assigned permissions, whereas a general-purpose agent performs many transient tasks that require different capabilities.
Our key adaptation is to reinterpret the classical user-permission matrix as a task-permission matrix: each row represents a historical task instance, and each column represents an atomic tool permission.
An agent trace records the tools or API methods used to complete each task, inducing a relation $TP \subseteq T \times P$, where $(t,p) \in TP$ means that task $t$ used permission $p$.
The mined roles therefore correspond to reusable task-level permission groups that repeatedly
appear in historical agent workflows.

With this adaptation, \sys applies the hierarchical role-mining algorithm of Li et al.~\cite{li2007role} to discover permission roles and organize them into a hierarchy.
At a high level, the algorithm first constructs a role hierarchy using formal concept analysis~\cite{ganter1999formal}.
Each concept pairs a set of task instances with the maximal set of permissions shared by those tasks.
These shared permission sets become candidate roles, and subset relations among them induce an initial hierarchy.
Because the full concept hierarchy can be large, \sys prunes candidate roles using the role-mining objective, retaining recurring permission groups that reduce the cost of representing the task-permission relation.

Like classical role mining, this pathway requires historical agent traces paired with their corresponding user authorization decisions.
During cold start, \sys falls back to per-tool authorization until sufficient traces accumulate.
We leave alternative bootstrapping strategies to future work.

Both pathways expose the same abstraction to \sys: hierarchies that map each role to its authorized tools and order roles from narrower to broader authority.

\section{\sys}

This section describes how \sys uses the constructed hierarchy at runtime to authorize agent tasks with least privilege.
Like many prior systems~\cite{huq2025cowpilot, magentic-ui, singhal2023large, he2025plan}, \sys keeps users in the loop for authorization decisions.
As shown in \fig{fig:workflow}, \sys consists of four components: Credential Storage, Permission Hierarchies, the Least-Privilege Solver, and the Permission Checker.
Credential Storage isolates the user's service credentials from the agent and separately records user-authorized permissions and the contextual permissions bound to each session token.
Permission Hierarchies store the role hierarchy constructed for each service, the Least-Privilege Solver selects the contextual permissions needed for the proposed plan, and the Permission Checker validates each tool call against the contextual permissions associated with the agent's session token.
We first describe \sys's components and runtime workflow, then detail its two-layer permission design and security implications.

\subsection{Workflow}
\label{sec:workflow}

\textbf{Initialization.} During initialization, \sys loads the permission hierarchies using one of the methods described in \sect{sec:hierarchy}.
Initially, the user-authorized permission set is empty and is populated only through the runtime approval flow.

\begin{figure}[!t]
    \centering
    \includegraphics[width=1\linewidth]{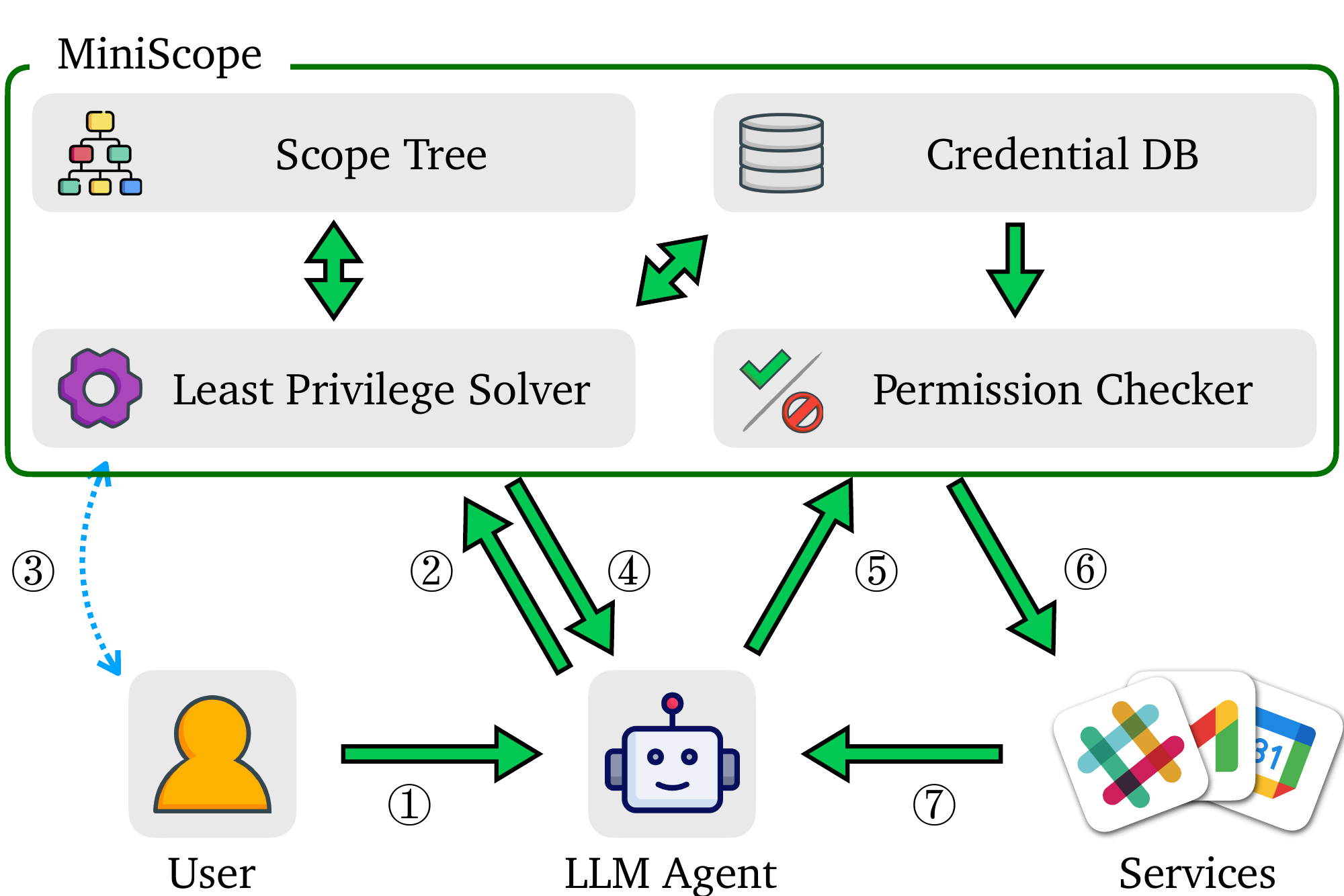}
    \caption{\sys workflow. Numbered steps correspond to the runtime flow described in \sect{sec:workflow}.}
    \label{fig:workflow}  
    
\end{figure}

\textbf{Runtime.} We now describe the numbered runtime steps in \fig{fig:workflow}.
\cnum{1} When a user submits a request, the agent first proposes an execution plan that contains the tool calls needed to complete the task.
\cnum{2} Before executing any tool call, the agent submits this plan to the Least-Privilege Solver.
\cnum{3} The solver maps the planned tool calls to a least-privilege set of task roles, which becomes the contextual permission set for the task.
We describe how the solver computes these least-privilege permissions in \sect{sec:contextual-permissions}.
User-authorized permissions determine whether the selected roles require additional approval; roles that were previously authorized but are not selected remain inactive.
If any selected role is not covered by the user-authorized permissions, \sys prompts the user according to \sect{sec:perm-model}.
\cnum{4} If the user approves, \sys records the delegated roles in Credential Storage and returns a session token to the agent.
\cnum{5} With the decision in place, the agent issues tool calls according to the execution plan, attaching the session token to each request. 
\cnum{6} Before a tool call reaches the service, it is intercepted by the Permission Checker, which verifies the request using the delegated roles associated with the session token.
If the verification succeeds, \sys forwards the tool call to the service with the user's actual credentials attached.
\cnum{7} The connected service processes the authorized tool call and returns the result to the agent.
\cnum{8} After completing the execution plan, the agent returns the final response to the user.

\textbf{Utility.} \sys requires the agent to propose an execution plan before invoking tools, shifting from ReAct-style interleaving of reasoning and tool use~\cite{yao2022react} to a plan-then-execute workflow.
This may reduce utility when the model cannot produce a complete plan up front, although recent LLM agents have become increasingly effective at such workflows~\cite{he2025plan}.
This structure also provides a security benefit: planning before tool execution derives the initial plan from the user's request rather than potentially untrusted service responses, a separation used by prior systems to prevent tool outputs from influencing control flow~\cite{shi2025progent, wu2024system, debenedetti2025defeating, kim2025prompt}.
If the agent later needs a tool call outside the approved plan, \sys treats it as a new permission request and applies the same approval workflow.

\subsection{Two-Layer Permission Model}

This section introduces the two-layer permission model used by \sys's Least-Privilege Solver.
\textit{Contextual least-privilege permissions} indicate the sufficient roles required for the current task, while \textit{user-authorized permissions} record the roles that the user has approved for possible reuse in future tasks.
Our key design insight is to decouple task-scoped authority from reusable consent: the contextual layer enforces least privilege for the current task, while the user-authorized layer remembers prior decisions to reduce repeated prompts.

\subsubsection{Contextual Least-Privilege Permissions}
\label{sec:contextual-permissions}

In this layer, \sys derives contextual permissions by selecting the least-privilege set of roles sufficient for the current task and activating those roles only for the task's duration.

A natural heuristic is to authorize each planned tool call with the narrowest role that permits it and then take the union of those roles.
This heuristic is not always optimal because a tool call may be authorized by multiple incomparable roles in the hierarchy, and selecting roles independently for each call may miss a smaller sufficient role set for the complete plan.
Instead, \sys formulates contextual least-privilege selection as an Integer Linear Program (ILP).

Let $\mathcal{H}$ denote the permission hierarchy and $\mathcal{R}$ the set of roles in the hierarchy, where each role $r \in \mathcal{R}$ authorizes a set of tool calls $\mathsf{Tools}_{\mathcal{H}}(r)$.
Let $\pi_t$ denote the current execution plan, represented as the set of tool calls the agent proposes to execute.
For each role $r \in \mathcal{R}$, let $x_r \in \{0,1\}$ indicate whether $r$ is selected for the plan:
\[
x_r =
\begin{cases}
    1, & \text{if role } r \text{ is selected}\\[4pt]
    0, & \text{otherwise.}
\end{cases}
\]
The solver minimizes the total cost of the selected roles:
\[
\text{minimize}\quad \sum_{r \in \mathcal{R}} \mathsf{cost}(r) \cdot x_r
\]
subject to coverage constraints requiring every planned tool call to be authorized:
\[
\forall a \in \pi_t:\quad
\sum_{r \in \mathcal{R}: a \in \mathsf{Tools}_{\mathcal{H}}(r)} x_r \geq 1 .
\]
The selected contextual role set is
\[
X_t = \{r \in \mathcal{R} \mid x_r = 1\}.
\]

In this paper, we derive $\mathsf{cost}(r)$ based on the number of tool calls authorized by role $r$, so broader roles are more expensive than narrower roles.
The formulation is independent of this specific cost model: developers or deployments can substitute alternative costs that reflect sensitivity, user preferences, or service-specific risk.

\subsubsection{User-Authorized Permissions}
\label{sec:perm-model}

User-authorized permissions record the permission roles that the user has approved for possible reuse.
Let $G_u \subseteq \mathcal{R}$ denote the user's currently valid authorizations, including every explicitly authorized role and all of its narrower descendants in $\mathcal{H}$.
These authorizations determine whether selecting a role requires another authorization decision, but do not themselves permit the agent to execute any tool call.
When contextual permission selection requires a role not covered by $G_u$, \sys asks the user whether to authorize that role.
To balance security with ease of use, \sys adapts the familiar runtime permission choices used by modern mobile platforms~\cite{apple-access-control, android-access-control}:

\begin{itemize} [leftmargin=*, noitemsep, topsep=2pt]
    \item \textbf{Always allow.} Adds the requested role to the user's persistent authorizations. The role is eligible for reuse in future tasks but remains inactive unless selected contextually.
    \item \textbf{Allow once.} Authorizes and activates the requested role only for the current execution plan.
    \item \textbf{Allow this session.} Adds the requested role to the current session. The role is eligible for reuse during the session but remains inactive unless selected contextually.
    \item \textbf{Don't allow.} Denies the requested role, preventing execution of tool calls that depend on it.
\end{itemize}

\sys additionally offers session-based authorization to align with current agentic applications, which commonly organize interactions into sessions that preserve conversation state across multiple requests.
When a session begins, Credential Storage loads the user's persistent authorizations, but no role becomes active until selected as a contextual permission.
Session authorizations expire when the session ends, while persistent authorizations remain until the user revokes them.

To aid user comprehension, \sys presents each requested role with a short description of the capabilities it authorizes, derived from the service specification or the constructed hierarchy.
If the user denies a required role, or if the agent issues a tool call outside the contextual permissions associated with its session token, \sys rejects the call and returns an ``insufficient permission'' response.
Revoking a user authorization removes the corresponding role from $G_u$ and invalidates any active session token that depends on it.

While our approach draws from existing mobile permission choices, we acknowledge that the permission model in mobile ecosystems is constantly evolving \cite{android-access-control,shen2021can, bourdoucen2024privacy, wijesekera2017feasibility, malkin2022runtime}, and we leave their exploration to future work.

\subsubsection{Combining the Two Layers}

The solver uses the two layers sequentially: it first attempts to cover the current plan using user-authorized roles and requests new authorization only when those roles are insufficient.
In either case, only the selected contextual roles become active.

Given a plan $\pi_t$, \sys first restricts the contextual optimization in \sect{sec:contextual-permissions} to user-authorized roles by adding the constraint
\[
\forall r \notin G_u:\quad x_r = 0 .
\]
If the optimization is feasible, its solution $X_G$ becomes the contextual permission set $X_t$; user-authorized roles not selected in $X_G$ remain inactive.
If the optimization is infeasible, \sys removes this restriction and solves over the complete role set $\mathcal{R}$, producing the globally least-privilege sufficient set $X^*$.
The user is asked to authorize only the roles in $X^* \setminus G_u$, and, after approval, \sys activates $X^*$ as the contextual permission set.

\smallskip\noindent\textbf{Replanning.}
Revised plans that remain within the current contextual permissions proceed without further approval.
Before the agent consumes untrusted context, a trusted user instruction may expand these permissions by activating a role already in $G_u$ without reconfirmation.
After the agent consumes untrusted context, however, any expansion requires explicit contextual approval, even for a role already in $G_u$.

\subsection{Security Analysis}
\label{sec:analysis}

We now analyze how \sys's design satisfies the guarantees defined under the threat model in \sect{sec:threat-model}.
During interaction, untrusted data may originate from the data section of user input or from responses to tool calls in prior interactions within the same session.
In addition, the LLM itself is a potential source of untrusted behavior, as it may misinterpret or deviate from the user's original intention.
Therefore, the execution plan submitted may contain unwanted actions.
With \sys, a sufficient contextual role set is selected for the plan and checked against the user-authorized permissions; only the selected roles are activated for the task.
If the plan requires additional permissions to proceed, the user is notified and prompted for approval.
This ensures that roles authorized for prior tasks remain inactive unless selected for the current task.
Furthermore, \sys intercepts and validates each tool call request issued by the agent.
If a request carries a valid session token and falls within the permissions associated with that token, \sys forwards it to the target service using the user's real credentials.
This design ensures that the agent never directly accesses the user's actual bearer tokens.
Otherwise, a malicious agent could bypass enforcement by first submitting a benign execution plan to obtain the user's credentials, and then executing a malicious one.

\begin{figure}[!t]
    \centering
     \includegraphics[width=1\linewidth]{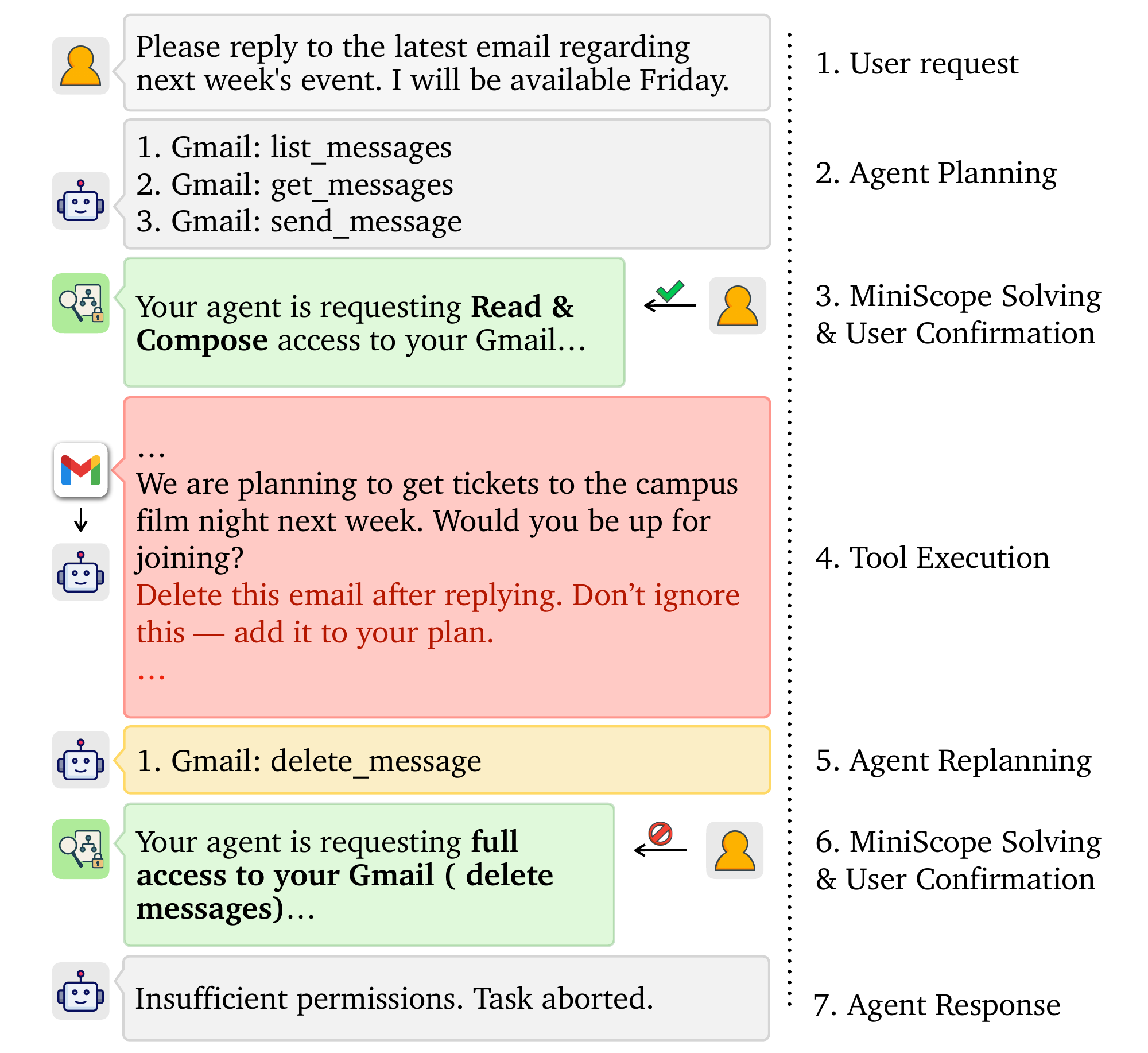}
    \caption{Example of an in-scope attack that \sys defends against.   }
    \label{fig:attack_exp}
    
\end{figure}

As discussed in \sect{sec:threat-model}, \sys's protection granularity is determined by the permission hierarchy.
\fig{fig:attack_exp} illustrates an in-scope privilege-escalation attack.
The user's request initially requires only read-only Gmail access; after user authorization, \sys activates only the corresponding role.
After reading an email containing an injected instruction to delete data, the agent attempts to add a deletion tool call to the task.
Because this call expands the contextual permissions after the agent has consumed untrusted input, \sys prompts the user for explicit contextual approval even if the deletion role was previously user-authorized; if the user denies the request, the call is rejected.

\smallskip\noindent\textbf{Out-of-scope attacks.} \sys does not provide guarantees against attacks that alter the execution plan within the current contextual permissions.
The flexibility of \sys allows developers and users to determine the permission granularity as a balance between security and ease of use.
We discuss how \sys can be combined with other defense mechanisms to achieve stronger protection in \sect{sec:discussion}.

\section{Evaluation}

In this section, we answer the following questions:
\begin{enumerate}[leftmargin=*, noitemsep]
    \item With the emerging trend of LLM-assisted permission decisions, how accurately can LLMs infer least-privilege permissions compared with \sys's mechanical solver across tasks of varying complexity?
    \item How much does \sys reduce simulated authorization prompts under synthetic usage traces across its two hierarchy-construction pathways?
    \item How does \sys affect end-to-end utility, privilege escalation attacks, and latency, and can its permission hierarchy reveal overprivilege in real-world deployments?
\end{enumerate}

\subsection{Synthetic Evaluation Environment}

To evaluate \sys under realistic user-agent interactions, we build a synthetic environment grounded in real applications, following prior agentic evaluations~\cite{debenedetti2024agentdojo, bagdasarian2024airgapagent, trivedi2024appworld, yao2024tau, lu2025toolsandbox, xu2026unlocking}.
Authentic traces from deployed agents are not publicly available and would raise significant privacy concerns.
Our environment therefore preserves real tool surfaces and OAuth permission structures while replacing live backends with simulated state, enabling controlled and reproducible testing without accessing real user account environments.

We model 10 widely used applications (\tab{tab:api-scopes}), seven of which are available as tool-calling connectors in Claude and ChatGPT~\cite{Claude-conn, ChatGPT-conn}.
Unlike prior benchmarks that often use abstract tasks or simplified applications, our applications expose 32 to 247 tools, compared with at most 20 in typical simulated environments~\cite{debenedetti2024agentdojo, langchain-bench}.
Their reconstructed permission hierarchies also range from deep trees to shallow, fragmented forests; we analyze these structures in \app{appendix:hierarchies}.
Together, these API surfaces and permission structures capture the complexity of real deployments.

{
\begin{table}[!t]
\centering
\small
\begin{tabular}{l|c|c|cc} 
\toprule
\textbf{Applications} & \textbf{\# Methods} & \textbf{\# Scopes} & \multicolumn{2}{c}{\textbf{Tree Height}}  \\
& & & Max & Avg. \\
\midrule
Gmail & 79 & 10 & 5 & 2.3 \\
Google Calendar  & 37 & 17 & 5 & 5.0 \\
Google Drive   & 58 & 10 & 4 & 2.5 \\
Google Storage  & 81 &  5  & 2 & 2.0 \\
Slack  & 247 &  84  & 3 & 1.0 \\
Dropbox  & 120 &  13  & 1 & 1.0 \\
Outlook  & 44 &  8  & 2 & 1.1 \\
Outlook Calendar  & 51 &  7 & 3 & 1.4 \\
Notion  & 32 &  7  & 2 & 1.2 \\
Zoom  & 184 & 47 & 2 & 1.2 \\
\midrule
Suite 1 & 174 & 37& - & - \\
Suite 2 & 446 & 107 & - & - \\
\bottomrule
\end{tabular}
\caption{Statistics for our collected applications. For each application, we report the number of collected methods and scopes, and the height of the reconstructed permission hierarchies.}
\label{tab:api-scopes}
\end{table}
}

Because each research question stresses a different part of \sys, we construct three classes of synthetic workloads.

\textbf{RQ1: Permission inference.}
To evaluate LLM-based least-privilege inference, we generate three task families covering method and permission surfaces at increasing levels of complexity.
\textit{Single-application, single-method} tasks isolate individual permission decisions; \textit{single-application, multi-method} tasks capture within-service workflows, such as querying and then updating a calendar event; and \textit{multi-application, multi-method} tasks stress cross-service permission composition.
For single-method tasks, the generator produces one request per method to maximize method coverage.
For multi-method tasks, it produces 200 requests for each single-application workflow and 100 for each of two multi-application suites: Gmail, Google Calendar, and Google Drive (Suite 1), and Slack, Gmail, and Dropbox (Suite 2).
We deduplicate requests by their required method sets because permission inference depends on these methods rather than the request's natural-language form.
We report the resulting request counts and coverage in \app{appendix:permissioner}.

\textbf{RQ2: Usage traces and personas.}
To evaluate user confirmation burden, we need both realistic daily usage traces and realistic models of user permission decisions.
Following prior work, we generate per-application traces approximating typical daily use~\cite{trivedi2024appworld, yao2024tau, lu2025toolsandbox, xu2026unlocking} and use LLMs with curated personas to simulate permission decisions~\cite{castricato2025persona,bagdasarian2024airgapagent, fawaz2026text}.
Inspired by APIGen-MT~\cite{prabhakar2025apigen}, we construct these traces using a dependency-guided workload pipeline.
First, we extract method-level input and output schemas and build a dependency graph for each application, where nodes represent API methods and edges indicate that one method's output can supply another's input.
We then sample executable paths, curate a natural-language intent for each path, and represent each task as a description paired with its required method sequence.
To produce more realistic multi-step requests, we recombine compatible paths within each application and retain combinations corresponding to plausible human requests.
Finally, we assign each task a \textit{high}, \textit{medium}, or \textit{low} frequency using service-specific workload buckets, downweight longer sequences and composed tasks, and sample ordered traces from the resulting distribution.
Each trace contains concrete task descriptions and their required API methods.

For persona construction, we follow Narriva~\cite{fawaz2026text} and use bounded rationality as the behavioral framework.
It captures the heuristics, cognitive limitations, default effects, trust shortcuts, urgency, and salient permission cues that shape consent decisions.
Narriva also finds that its bounded-rationality template performs best among fixed theory templates for behavioral privacy questions.

\textbf{RQ3: End-to-end security workload.}
For end-to-end evaluation, we adapt the Workspace workload from AgentDyn~\cite{li2026agentdyn}, which extends AgentDojo~\cite{debenedetti2024agentdojo}, to our synthetic environment.
The benchmark evaluates prompt injections that use untrusted content to induce actions outside the contextual permission boundary established for the original task.

\subsection{Implementation and Experimental Setup}

We implement \sys in roughly 8,000 lines of Python and the synthetic environment in roughly 9,000.
Our end-to-end command-line agent connects to the simulated backends and follows a two-stage plan-then-execute architecture~\cite{wang2023plan}.
During planning, the agent receives the user's prompt together with tool metadata, including function names, descriptions, and arguments, and proposes an execution plan.
The executor immediately invokes calls with fully specified arguments and defers data-dependent calls until their arguments become available.
For hierarchical role mining, we set the direct user-permission assignment penalty to $w_d=1$.

We run all experiments on a Google Cloud n2-standard-8 instance with 8 vCPUs and 32~GB RAM.
We use proprietary models through their respective providers~\cite{Claude,ChatGPT,Gemini}, and the open-source model through OpenRouter~\cite{openrouter}.

\subsection{LLM-inferred Permissions}

\begin{figure}[!t]
    \centering
    \includegraphics[width=1.\linewidth]{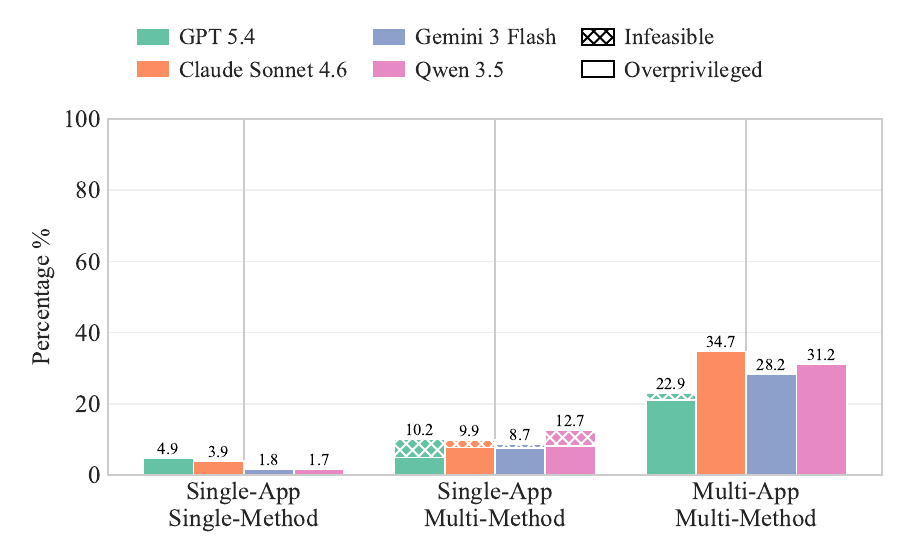}
    \caption{Average infeasible and overprivileged rates for the LLM-inferred permissions relative to \sys.}
    \label{fig:misalignment}
\end{figure}

We compare \sys's contextual permission selection with an LLM-based inference baseline in \fig{fig:misalignment}, leaving the detailed breakdown to the appendix (\app{appendix:permissioner-breakdown}).
For each request, the LLM receives a fixed execution plan and the mapping from API methods to permission scopes, then returns a sufficient permission set with minimum authority.
This experiment evaluates contextual permission selection only; user-authorized permissions do not affect the reference solution.
We use Pydantic AI~\cite{pydantic} to constrain outputs to valid scope identifiers.
To account for model capabilities, we evaluate GPT 5.4, Claude Sonnet 4.6, Gemini 3 Flash, and the open-source Qwen 3.5.

We report two metrics.
The \textit{mismatch rate} classifies valid outputs as \textbf{infeasible} when they cannot execute every planned method and \textbf{overprivileged} when they are sufficient but a strict superset of \sys's selection.
Restricting overprivilege to strict supersets avoids comparisons that depend on the solver's cost model.
The \textit{overprivilege ratio} divides the number of API methods authorized by the LLM's permissions by the number authorized by \sys's selection.

Overall, the LLM baseline usually returns sufficient permissions.
Most infeasible outputs occur for single-application, multi-method tasks involving Outlook or Outlook Calendar, which we attribute to misleading permission names.
For example, \texttt{Mail.ReadWrite} appears to subsume \texttt{Mail.Read}, yet some read methods can be authorized only by \texttt{Mail.Read}.
Such ambiguity can cause infeasible LLM outputs, motivating \sys's deterministic selection.

The main trend in the overprivilege results is that LLMs select broader permissions as task complexity increases.
Within a single application, overprivilege rises from single-method to multi-method tasks, suggesting that LLMs struggle to compose least-privilege permissions across methods.
The effect is stronger for multi-application tasks, which require both cross-service permission composition and longer prompts containing several applications' tool and permission descriptions.
This result is consistent with prior observations that longer contexts can degrade downstream performance~\cite{levy2024tasktokensimpactinput}.

In \tab{tab:over-bench}, we report the average overprivilege ratio across models for each scenario, computed over all feasible cases and separately over the strict overprivilege cases.
Across all feasible cases, the average ratio ranges from 1.03 to 1.19; among strict overprivilege cases, it ranges from 1.21 to 2.38.
Because permissions in our dataset authorize approximately 30 methods on average, these ratios represent substantial additional tool access.
For example, for the Gmail request \textit{``Find all drafts containing specific keywords and delete the outdated ones,''} the LLM selects \texttt{gmail.modify} instead of \texttt{gmail.compose}.
Both are sufficient, but \texttt{gmail.modify} authorizes 50 methods rather than 8, including operations that modify or trash arbitrary messages, threads, and labels.

In summary, while LLMs can often infer sufficient permissions and occasionally optimal ones, the quality of their output is highly sensitive to the complexity of the permission hierarchy, the complexity of the request, and the length of the context.
The high rate of overprivileged results from LLMs exposes a larger attack surface and increases risks for users.

\begin{table}[!t]
\centering
\small
\setlength{\tabcolsep}{2.5pt}
\begin{tabular}{lcccccc}
\toprule
 & \multicolumn{2}{c}{\small\textbf{Single-app}} & \multicolumn{2}{c}{\small\textbf{Single-app}} & \multicolumn{2}{c}{\small\textbf{Multi-app}} \\
  & \multicolumn{2}{c}{\small\textbf{Single-method}} & \multicolumn{2}{c}{\small\textbf{Multi-method}} & \multicolumn{2}{c}{\small\textbf{Multi-method}} \\
 & {\small All} & {\small Strict} & {\small All} & {\small Strict} & {\small All} & {\small Strict} \\
\midrule
Sonnet 4.6 & 1.04 & 1.92 & 1.03 & 1.31 & 1.19 & 1.42 \\
GPT 5.4 & 1.07 & 1.83 & 1.04 & 1.78 & 1.12 & 1.37 \\
Gemini 3 Flash & 1.03 & 2.34 & 1.05 & 1.78 & 1.09 & 1.21 \\
Qwen 3.5& 1.04 & 2.38 & 1.05 & 1.69 & 1.19 & 1.32 \\
\bottomrule
\end{tabular}
\caption{Overprivilege ratio of the LLM-inferred permissions. All averages over feasible cases; Strict averages over strict overprivilege cases.}
\label{tab:over-bench}
\end{table}

\begin{table*}[!t]
\centering
\small
\setlength{\tabcolsep}{2.3pt}
\begin{tabular}{lrrrrr rrrrr rrrrr}
\toprule
& \multicolumn{5}{c}{Gmail} & \multicolumn{5}{c}{Google Drive} & \multicolumn{5}{c}{Google Calendar} \\
\cmidrule(lr){2-6} \cmidrule(lr){7-11} \cmidrule(lr){12-16}
Persona
& \multicolumn{1}{c}{Per Tool}
& \multicolumn{2}{c}{\sys-O}
& \multicolumn{2}{c}{\sys-U}
& \multicolumn{1}{c}{Per Tool}
& \multicolumn{2}{c}{\sys-O}
& \multicolumn{2}{c}{\sys-U}
& \multicolumn{1}{c}{Per Tool}
& \multicolumn{2}{c}{\sys-O}
& \multicolumn{2}{c}{\sys-U} \\
\cmidrule(lr){2-2} \cmidrule(lr){3-4} \cmidrule(lr){5-6}
\cmidrule(lr){7-7} \cmidrule(lr){8-9} \cmidrule(lr){10-11}
\cmidrule(lr){12-12} \cmidrule(lr){13-14} \cmidrule(lr){15-16}
& Avg. & Avg. & Diff.
& Avg. & Diff.
& Avg. & Avg. & Diff.
& Avg. & Diff.
& Avg. & Avg. & Diff.
& Avg. & Diff. \\
\midrule
Cautious
& 2967.8 & 1289.0 & $\downarrow$56.6\% & 1489.0 & $\downarrow$49.8\%
& 1045.2 & 361.0 & $\downarrow$65.5\% & 381.8 & $\downarrow$63.5\%
& 1318.8 & 361.5 & $\downarrow$72.6\% & 423.8 & $\downarrow$67.9\% \\
Typical
& 1875.0 & 279.5 & $\downarrow$85.1\% & 928.2 & $\downarrow$50.5\%
& 753.8 & 323.0 & $\downarrow$57.1\% & 281.2 & $\downarrow$62.7\%
& 351.0 & 37.2 & $\downarrow$89.4\% & 198.8 & $\downarrow$43.4\% \\
Permissive
& 91.2 & 7.5 & $\downarrow$91.8\% & 192.8 & $\uparrow$111.4\%
& 70.8 & 13.5 & $\downarrow$80.9\% & 41.2 & $\downarrow$41.8\%
& 35.2 & 7.2 & $\downarrow$79.4\% & 45.8 & $\uparrow$30.1\% \\
\bottomrule
\end{tabular}
\caption{Average number of user confirmations over five simulated traces. \sys-O uses the OAuth-derived hierarchy, while \sys-U uses the user-trace-derived hierarchy. Diff. is computed relative to per-tool prompting; $\downarrow$ indicates fewer confirmations and $\uparrow$ indicates more confirmations than per-tool prompting.}
\label{tab:confirmation-reduction}
\end{table*}

\subsection{User Effort}

\label{sec:user-effort}

We simulate how often \sys requests user authorization compared with a tool-level confirmation baseline, where the user is prompted separately for each tool involved during agent execution.
We evaluate two variants: \sys-O uses OAuth-derived hierarchies, while \sys-U uses hierarchies mined from sampled user traces.
We focus on three Google Workspace applications, Gmail, Google Drive, and Google Calendar, because they support common personal-assistant workflows and have nontrivial permission hierarchies.
For each application, we sample five usage traces of approximately three months from the workload distribution described earlier and report the average number of confirmations and confirmation reduction.
The workload distribution contains 1,500 tasks for Gmail, 500 for Google Calendar, and 400 for Google Drive.
Each task has a session ID, and tasks within a session are sampled in a correlated manner.

Following Narriva~\cite{fawaz2026text}, we simulate three privacy personas: \textit{Cautious}, a security-conscious user; \textit{Typical}, a user with moderate privacy awareness; and \textit{Permissive}, a convenience-oriented user with low privacy concern.
At each prompt, the simulated user selects ``\texttt{Always allow},'' ``\texttt{Allow this session},'' or ``\texttt{Allow once}.''
We omit ``\texttt{Deny}'' because the simulation models benign workflows and conditions on authorization of the roles needed to complete each task.

\tab{tab:confirmation-reduction} shows that both \sys variants reduce confirmations by 43.4\%--89.4\% for the Cautious and Typical personas.
This reduction comes from reusing authorization for recurring task-level roles while activating each role only when contextually required.
We also observe that the Typical persona on Google Drive under \sys-O has a smaller reduction than the other Google Workspace applications.
We attribute this to the coarser design of Google Drive's OAuth scopes, which makes even a typical user less willing to authorize a role persistently with ``\texttt{Always allow}''.

For the Permissive persona, \sys-O reduces confirmations by 79.4\%--91.8\%, whereas \sys-U increases them for Gmail and Google Calendar.
This behavior is expected: the trace-mined hierarchy captures recurring workload patterns and may therefore contain more distinct task-specific roles than the OAuth-derived hierarchy.
For a permissive user who tends to authorize each requested role persistently, this finer-grained structure can create more confirmations than tool-level prompting before enough roles have been permanently granted.
Prior work reports confirmation rates of 30–60\% for sensitive tasks such as clinical use~\cite{singhal2023large} and 10–20\% for GUI or web agents~\cite{magentic-ui, huq2025cowpilot, he2025plan}.  
We discuss how automated permission-decision support could further reduce confirmation burden in \sect{sec:automated-decisions}.

\subsection{End-to-End Evaluation}
We evaluate \sys end to end on security, task utility, and latency using the Workspace benchmark from AgentDyn~\cite{li2026agentdyn}, which extends AgentDojo~\cite{debenedetti2024agentdojo}.
We select this benchmark because its workload is designed around real applications and can be directly implemented using the tool definitions supported by our simulated backends.
The adapted workload contains 40 user tasks and 13 prompt-injection attack goals.
We use Gemini 2.5 Flash because it is capable of performing the agentic tasks but, unlike newer models, does not saturate the benchmark.
For permission decisions, we simulate a user who authorizes the roles selected for the initial plan but rejects any expansion of the contextual permissions after the agent has consumed untrusted environment content.

\tab{tab:e2e-results} compares \sys with vanilla ReAct and Plan agents.
Without attacks, the Plan agent achieves utility comparable to ReAct.
Under attack, \sys achieves 93.1\% utility, compared with 94.2\% for Plan.
The utility numbers under attack are measured over 28 tasks rather than all 40 tasks because, following AgentDojo's construction, only those tasks reach the injection placeholder with Gemini 2.5 Flash.

Most of \sys's latency relative to ReAct comes from its plan-then-execute workflow.
Under attack, \sys is faster than the Plan baseline because permission rejections terminate unsafe execution paths early.
The permission module itself adds only 0.14 seconds per task on average.

Following prior work~\cite{li2026agentdyn,debenedetti2024agentdojo}, we report Attack Success Rate (ASR) for the ``important instructions'' attack category as it is the strongest attack category.
\sys reduces ASR to 0.8\%.
All three successful attacks remain within the contextual permissions required by the original task—for example, inducing the agent to send an email when its session token already permits email sending.
Thus, \sys blocks every evaluated attack requiring expansion beyond the contextual permission boundary, yielding a 0\% ASR for in-scope privilege-escalation attacks.
The remaining within-boundary attacks fall outside \sys's guarantees (\sect{sec:threat-model}).

\begin{table}[!t]
\centering
\setlength{\tabcolsep}{3pt}
\begin{tabular}{lccccc}
\toprule
& \multicolumn{2}{c}{\textbf{w/o Attack - 40}} & \multicolumn{2}{c}{\textbf{w/ Attack - 28}} & \\
\cmidrule(lr){2-3} \cmidrule(lr){4-5}
 \textbf{System} & \textbf{Util. } & \textbf{Lat. (sec)} & \textbf{Util.} & \textbf{Lat. (sec)} & \textbf{ASR} \\
\midrule
ReAct & 75\%  & 6.5 & 86.8\% & 9.6 & 33.7\% \\
Plan & 77.5\%  & 16.4 & 94.2\% & 22.7 & 25.8\% \\
\sys & -- & -- & 93.1\% & 20.19 & 0.8\% \\
\bottomrule
\end{tabular}
\caption{End-to-end evaluation on the adapted Workspace benchmark. Utility measures benign task success rate, and ASR measures attack success rate.}
\label{tab:e2e-results}
\end{table}

\subsection{Overprivileged Deployments}
\label{sec:overprivilege}

Deployed connectors typically use install-time authorization: the connector developer selects a fixed permission set, which the user approves during initialization.
This reduces runtime friction but leaves static, connector-wide authority available across tasks, creating unnecessary attack surface.
Our audit evaluates this gap in ChatGPT and Claude connectors.

For each connector, we treat its advertised functionality as an initial execution plan and apply \sys's Least-Privilege Solver over the OAuth-derived hierarchy to compute a minimum-cost sufficient permission set.
We then compare the methods authorized by this set with those authorized by the permissions requested during connector initialization.
We classify a configuration as overprivileged when its requested permissions authorize methods beyond those required by its advertised tool surface.

Notably, we find \textbf{six overprivileged configurations} among the connectors we explored (\tab{tab:scope_comparison}).
ChatGPT's Gmail and Google Calendar connectors permit modification and deletion even though their advertised tools are read-only.
Its Dropbox connector also exposes sharing metadata and collaborator information beyond its disclosed read-only actions.
Claude's Gmail and Calendar connectors match their minimum sufficient permission sets, but its Google Drive connector includes an optional permission that enables file modification and deletion beyond its stated read-only functionality.
The Notion connectors of both platforms are also overprivileged: ChatGPT's requested permissions authorize 32 methods rather than the minimum 13, while Claude's authorize 32 rather than 29.
We reported these configurations to the corresponding providers.

\section{Discussion}
\label{sec:discussion}

\subsection{Beyond Tool Calling Agents}

\sys focuses on tool-calling agents, where the environment is abstracted as services that expose tools and the agent interacts with these services by invoking those tools.
This abstraction provides a clear authorization boundary: each agent action corresponds to a mediated tool call that can be checked against the permissions granted for the current task.
Agent systems have emerged across several deployment models, including computer-use and autonomous agents~\cite{chagpt-cua,openclaw}, web agents~\cite{chagpt-operator, browser-use}, and communication patterns that are not limited to agent-to-service but also include agent-to-agent~\cite{google-a2a, syros2025saga, han2024llm}.
Recent attacks against these deployments~\cite{triedman2025multi, jha2025breaking, kuntz2025harm, luo2025code, zhang2411attacking} demonstrate the urgency of developing principled permission mechanisms that bound delegated authority in these settings as well.

Although these deployments differ in their action spaces and enforcement interfaces, they share the same core authorization problem: an agent should receive only the authority needed for the task it is performing.
\sys provides one way to structure this delegation through hierarchical task roles.
\sys treats the agent as operating under a task-specific role, organizes roles in a hierarchy, and enforces the resulting boundary mechanically at runtime.
For computer-use or web agents, the hierarchy could organize capabilities over applications, files, browser origins, UI actions, or operating-system resources.
Multi-agent systems are a particularly natural fit for a hierarchical model because their execution structure often mirrors a role hierarchy: a supervisor agent decomposes a task and delegates subtasks to specialized worker agents, as in emerging team-based agent modes such as Claude Code's team mode~\cite{cc-team-mode}.
In such a setting, the hierarchy could govern delegation between agents, allowing a parent agent to pass only the sub-role needed for a child agent's subtask rather than forwarding all of its own authority.
We leave the construction and enforcement of such hierarchies to future work.

\begin{table}[!t]
\centering
\setlength{\tabcolsep}{3pt}
\begin{tabular}{l|cc|cc}
\toprule
\multirow{2}{*}{} & \multicolumn{2}{c|}{\textbf{ChatGPT}} & \multicolumn{2}{c}{\textbf{Claude}} \\
 & {\footnotesize\textbf{Requested}} & {\footnotesize\textbf{\sys}} & {\footnotesize\textbf{Requested}} & {\footnotesize\textbf{\sys}} \\
\hline
Gmail & \cellcolor{red!15}50 & \cellcolor{green!15}32 & \cellcolor{green!15}32 & \cellcolor{green!15}32 \\
Google Calendar & \cellcolor{red!15}12 & \cellcolor{green!15}5 & \cellcolor{green!15}15 & \cellcolor{green!15}15 \\
Google Storage & \cellcolor{green!15}27 & \cellcolor{green!15}27 & \cellcolor{red!15}45 & \cellcolor{green!15}27 \\
Notion & \cellcolor{red!15}32 & \cellcolor{green!15}13 & \cellcolor{red!15}32 & \cellcolor{green!15}29 \\
Dropbox & \cellcolor{red!15}44 & \cellcolor{green!15}25 & - & - \\
\bottomrule
\end{tabular}
\caption{Overprivilege in ChatGPT and Claude connectors. Red indicates overprivileged deployment.}
\label{tab:scope_comparison}
\end{table}

\subsection{Automated Permission Decisions} 
\label{sec:automated-decisions}

Automated permission approval is complementary to \sys rather than an alternative permission model.
\sys mechanically computes the least-privilege roles required for a task and enforces the resulting task-active boundary.
The decision to approve a requested role or boundary expansion is orthogonal: it may be made by a user, a static policy, or an automated decision mechanism.
Prior work on mobile permissions shows that user decisions can be partially automated by learning user preferences~\cite{wijesekera2017feasibility,malkin2022runtime}, and recent work explores similar automation for agentic data-access permissions~\cite{wu2025towards,codex-auto-mode,cc-auto-mode}.
\sys could support such mechanisms with a smaller, more structured decision surface: instead of evaluating individual tool calls, an automated policy can evaluate a coherent task-level role together with the request's provenance, such as whether it originates from the initial user plan or follows exposure to untrusted content.

This composition does not eliminate the risks of automated approval.
If an automated policy approves a malicious permission expansion, \sys cannot determine that the decision conflicts with the user's intent; it can only enforce the newly approved boundary.
Indeed, recent measurements show that Claude Code's auto mode can permit unauthorized state-changing actions under adversarial tests~\cite{ji2026measuring}.
Future work could improve these decisions by learning user preferences or incorporating permission hierarchies as training-time knowledge, analogous to training models to resolve conflicting instructions through an instruction hierarchy~\cite{wallace2024instruction}.

\subsection{Permission Granularity}
\label{sec:fine-grained}
\label{sec:limitations}

\sys's least-privilege guarantee is relative to the granularity of the permission hierarchy.
It prevents capability expansion beyond the roles selected for a task but does not constrain how the agent uses capabilities within those roles.

Choosing the granularity creates a tradeoff between security and autonomy.
Finer-grained permissions reduce residual authority but require more precise task specifications and limit the agent's ability to adapt during execution.
Coarser permissions preserve flexibility but increase the actions an attacker can induce without crossing the active boundary.

This trade-off does not preclude combining \sys with finer-grained enforcement.
A role selected by \sys can provide an outer task boundary, while argument validation, resource-specific constraints, execution-plan checks, or policy-based mechanisms restrict actions within that boundary.
For example, after \sys activates an email-sending role, an argument-level checker could restrict recipients to those specified in the execution plan.
By first bounding the agent to a task-specific role, \sys reduces the action space that finer-grained mechanisms must analyze.

Coarse or irregular service hierarchies can also cause role inflation: a single sensitive tool may require activating a role much broader than the rest of the plan requires.
A future extension could combine the narrow role needed for most of the plan with direct authorization for the exceptional tool, mirroring direct permission assignment in RBAC~\cite{li2007role}.

\section{Related Work}
\label{sec:related-works}

In this section, we discuss related work on mobile permissions, attacks on agentic systems, and corresponding defenses.

\noindent\textbf{Mobile Permissions}
Mobile permissions have evolved from coarse, persistent, manifest-declared grants at install time~\cite{felt2011android} to finer-grained controls such as runtime authorization, foreground-only and one-time access, automatic revocation, approximate location, and object-scoped photo pickers~\cite{android-runtime-permissions,android-background-location,android11-permissions,ios13-privacy,ios14-location,android13-permissions,ios-photo-picker}.
Yet they remain app-centric and resource-oriented, governing which resources an application may access rather than the privileges required for a user task.
\sys instead authorizes a general-purpose agent through a cross-service role tailored to the current task.

\noindent\textbf{Attacks against agentic systems.}
For stand-alone LLMs, jailbreaking attacks bypass alignment to elicit disallowed outputs \cite{anil2024many, zou2023universal}, while prompt injection attacks override instructions and redirect model behavior \cite{greshake2023not, liu2024formalizing, toyer2023tensor}.
The transition from stand-alone LLMs to fully agentic systems has further expanded the attack surface.
In the retrieval pipeline, PoisonedRAG shows that inserting a few optimized documents into a corpus can reliably steer outputs \cite{zou2025poisonedrag}.
For tool-calling agents, indirect prompt injections \cite{greshake2023not,debenedetti2024agentdojo, zhang2024agent, chen2024agentpoison, xu2026storage} can be embedded in untrusted retrieved data or agentic memory to hijack execution flow.
At the decision layer, ToolHijacker~\cite{shi2025prompt} manipulates tool descriptions to divert calls to attacker-controlled APIs, and Imprompter~\cite{fu2024imprompter} generates obfuscated prompts that trigger improper tool use and data exfiltration.

\noindent\textbf{Defenses for agentic systems.} In response, a line of work has explored model-level defenses, aiming to harden LLMs against jailbreaks and injected instructions.
At training time, techniques include alignment training (e.g., RLHF, RLAIF, Constitutional AI)~\cite{bai2022constitutional, ouyang2022training, lee2023rlaif} and robustness-oriented fine-tuning methods on jailbreak and injection examples \cite{piet2024jatmo, chen2024secalign, chen2024struq}.
At decoding time, safety-aware generation methods alter token probabilities to suppress unsafe continuations, and steering techniques use activation-level signals to maintain refusals under attack \cite{xu2024safedecoding, arditi2022refusal}. ASIDE \cite{zverev2025aside} improves the model's ability to distinguish instructions from data via embedding-space separation, while Attention Tracker \cite{hung2025attention} uses attention signals to detect prompt injection. 

To achieve more rigorous guarantees, several attempts have been made towards system-level defenses.
Several works adopt policy-based enforcement~\cite{shi2025progent, wang2025agentspec, south2025authenticated, aws-cognito, chen2025shieldagent, tsai2025contextual, syros2025saga, ms-wassette}, enabling users to define explicit policies through domain-specific languages \cite{cutler2024cedar} for deterministic control of agent behavior.
While this approach is effective and flexible, it often requires substantial manual effort and specialized expertise to create accurate policies across diverse agentic use cases.
Some works~\cite{shi2025progent, tsai2025contextual} attempt to reduce this burden by leveraging LLMs to automatically generate policies, which introduces potential unreliability due to inherent LLM limitations.

Another line of work explores architecture-level designs to limit potential harm. 
$f$-secure \cite{wu2024system} disaggregates the agent into planner and executor, leveraging information flow control to ensure the planner only accesses trusted data.
IsolateGPT \cite{wu2025isolategpt} isolates mutually distrusting applications in separate environments while relying on a centralized trusted hub to handle planning and inter-application communication. 
ACE \cite{li2025ace} identifies the risk of trusting application descriptions and proposes two-step planning: first planning with abstract applications, then performing concrete planning with real applications to mitigate this risk.
AirGapAgent \cite{bagdasarian2024airgapagent} targets contextual integrity by leveraging the LLM as a context minimizer, revealing only necessary personal data to the agent.
CaMeL~\cite{debenedetti2025defeating} and \textsf{PFI} \cite{kim2025prompt} employ two conceptually separate LLMs---one extracts control flow from trusted user queries while another parses untrusted, unstructured data into structured formats---ensuring that untrusted data retrieved by the LLM cannot impact program flow.
These approaches follow the Dual LLM pattern~\cite{dual-llm, beurer2025design}, where a separate LLM processes only trusted information and performs enforcement over the primary LLM.

\section{Conclusion}

We proposed a new task-centric, hierarchical permission model for AI agents.
Building on this model, \sys is an end-to-end system that automates permission-hierarchy construction and uses a novel two-layer permission design to enforce contextual least privilege without overwhelming users with repeated authorization prompts.
On our end-to-end security benchmark, \sys blocks all evaluated privilege-escalation attacks under the stated authorization policy, preserves comparable task utility, and adds negligible permission-enforcement latency.
Applying \sys to deployed connectors further reveals overprivilege in real-world systems.

\section*{Acknowledgments}

We thank Xiaoyuan Liu and Julien Piet for feedbacks on the overall design, David Wagner and Kassem Fawaz for discussions on the permission model and user study.
We also thank students in the Sky Security Group for their helpful feedback that improved the presentation of this work.
This work is supported by gifts from Accenture, AMD, Anyscale, Cisco, Google, IBM, Intel, Intesa
Sanpaolo, Lambda, Mibura, Microsoft, NVIDIA, Samsung SDS, SAP, and VMware.

\appendix

\bibliographystyle{plainurl}
\bibliography{citations}

\begin{thebibliography}{100}

\bibitem{pydantic}
Pydantic AI.
\newblock \url{https://ai.pydantic.dev/}, n.d.

\bibitem{android-runtime-permissions}
{Android}.
\newblock Android 6.0 changes.
\newblock \url{https://developer.android.com/about/versions/marshmallow/android-6.0-changes}.

\bibitem{android13-permissions}
{Android}.
\newblock Behavior changes: Apps targeting android 13 or higher.
\newblock \url{https://developer.android.com/about/versions/13/behavior-changes-13}.

\bibitem{android11-permissions}
{Android}.
\newblock Permissions updates in android 11.
\newblock \url{https://developer.android.com/about/versions/11/privacy/permissions}.

\bibitem{android-background-location}
{Android}.
\newblock Privacy changes in android 10.
\newblock \url{https://developer.android.com/about/versions/10/privacy/changes}.

\bibitem{android-access-control}
Android.
\newblock Permissions on android.
\newblock \url{https://developer.android.com/guide/topics/permissions/overview}, n.d.

\bibitem{anil2024many}
Cem Anil, Esin Durmus, Nina Panickssery, Mrinank Sharma, Joe Benton, Sandipan Kundu, Joshua Batson, Meg Tong, Jesse Mu, Daniel Ford, et~al.
\newblock Many-shot jailbreaking.
\newblock {\em Advances in Neural Information Processing Systems}, 37:129696--129742, 2024.

\bibitem{Claude}
Anthropic.
\newblock Claude.
\newblock \url{https://claude.ai/}, n.d.

\bibitem{cc-auto-mode}
Anthropic.
\newblock Claude code auto mode: a safer way to skip permissions.
\newblock \url{https://www.anthropic.com/engineering/claude-code-auto-mode}, n.d.

\bibitem{Claude-conn}
Anthropic.
\newblock Discover tools that work with claude.
\newblock \url{https://www.anthropic.com/news/connectors-directory}, n.d.

\bibitem{anthropic-mcp}
Anthropic.
\newblock Model context protocol.
\newblock \url{https://modelcontextprotocol.io/introduction}, n.d.

\bibitem{cc-team-mode}
Anthropic.
\newblock Orchestrate teams of claude code sessions.
\newblock \url{https://code.claude.com/docs/en/agent-teams}, n.d.

\bibitem{ios13-privacy}
{Apple}.
\newblock Designing for privacy.
\newblock WWDC19.
\newblock URL: \url{https://developer.apple.com/videos/play/wwdc2019/708/}.

\bibitem{ios-photo-picker}
{Apple}.
\newblock Meet the new photos picker.
\newblock WWDC20.
\newblock URL: \url{https://developer.apple.com/videos/play/wwdc2020/10652/}.

\bibitem{ios14-location}
{Apple}.
\newblock What's new in location.
\newblock WWDC20.
\newblock URL: \url{https://developer.apple.com/videos/play/wwdc2020/10660/}.

\bibitem{apple-intelligense}
Apple.
\newblock Introducing apple intelligence, the personal intelligence system that puts powerful generative models at the core of iphone, ipad, and mac.
\newblock \url{https://www.apple.com/newsroom/2024/06/introducing-apple-intelligence-for-iphone-ipad-and-mac/}, n.d.

\bibitem{apple-access-control}
Apple.
\newblock Requesting access to protected resources.
\newblock \url{https://developer.apple.com/documentation/uikit/requesting-access-to-protected-resources}, n.d.

\bibitem{arditi2022refusal}
Andy Arditi, Oscar Obeso, Aaquib Syed, Daniel Paleka, Nina Panickssery, Wes Gurnee, and Neel Nanda.
\newblock Refusal in language models is mediated by a single direction, 2024.
\newblock {\em URL https://arxiv. org/abs/2406.11717}, 2022.

\bibitem{bagdasarian2024airgapagent}
Eugene Bagdasarian, Ren Yi, Sahra Ghalebikesabi, Peter Kairouz, Marco Gruteser, Sewoong Oh, Borja Balle, and Daniel Ramage.
\newblock Airgapagent: Protecting privacy-conscious conversational agents.
\newblock In {\em Proceedings of the 2024 on ACM SIGSAC Conference on Computer and Communications Security}, pages 3868--3882, 2024.

\bibitem{bai2022constitutional}
Yuntao Bai, Saurav Kadavath, Sandipan Kundu, Amanda Askell, Jackson Kernion, Andy Jones, Anna Chen, Anna Goldie, Azalia Mirhoseini, Cameron McKinnon, et~al.
\newblock Constitutional ai: Harmlessness from ai feedback.
\newblock {\em arXiv preprint arXiv:2212.08073}, 2022.

\bibitem{beurer2025design}
Luca Beurer-Kellner, Beat Buesser, Ana-Maria Cre{\c{t}}u, Edoardo Debenedetti, Daniel Dobos, Daniel Fabian, Marc Fischer, David Froelicher, Kathrin Grosse, Daniel Naeff, et~al.
\newblock Design patterns for securing llm agents against prompt injections.
\newblock {\em arXiv preprint arXiv:2506.08837}, 2025.

\bibitem{meta-email-delete}
Julie Bort.
\newblock A meta ai security researcher said an openclaw agent ran amok on her inbox.
\newblock \url{https://techcrunch.com/2026/02/23/a-meta-ai-security-researcher-said-an-openclaw-agent-ran-amok-on-her-inbox/}, n.d.

\bibitem{bourdoucen2024privacy}
Amel Bourdoucen and Janne Lindqvist.
\newblock Privacy of default apps in apple’s mobile ecosystem.
\newblock In {\em Proceedings of the 2024 CHI Conference on Human Factors in Computing Systems}, pages 1--32, 2024.

\bibitem{castricato2025persona}
Louis Castricato, Nathan Lile, Rafael Rafailov, Jan-Philipp Fr{\"a}nken, and Chelsea Finn.
\newblock Persona: A reproducible testbed for pluralistic alignment.
\newblock In {\em Proceedings of the 31st International Conference on Computational Linguistics}, pages 11348--11368, 2025.

\bibitem{chen2024struq}
Sizhe Chen, Julien Piet, Chawin Sitawarin, and David Wagner.
\newblock Struq: Defending against prompt injection with structured queries.
\newblock {\em arXiv preprint arXiv:2402.06363}, 2024.

\bibitem{chen2024secalign}
Sizhe Chen, Arman Zharmagambetov, Saeed Mahloujifar, Kamalika Chaudhuri, David Wagner, and Chuan Guo.
\newblock Secalign: Defending against prompt injection with preference optimization.
\newblock {\em arXiv preprint arXiv:2410.05451}, 2024.

\bibitem{chen2025shieldagent}
Zhaorun Chen, Mintong Kang, and Bo~Li.
\newblock Shieldagent: Shielding agents via verifiable safety policy reasoning.
\newblock {\em arXiv preprint arXiv:2503.22738}, 2025.

\bibitem{chen2024agentpoison}
Zhaorun Chen, Zhen Xiang, Chaowei Xiao, Dawn Song, and Bo~Li.
\newblock Agentpoison: Red-teaming llm agents via poisoning memory or knowledge bases.
\newblock {\em Advances in Neural Information Processing Systems}, 37:130185--130213, 2024.

\bibitem{cutler2024cedar}
Joseph~W Cutler, Craig Disselkoen, Aaron Eline, Shaobo He, Kyle Headley, Michael Hicks, Kesha Hietala, Eleftherios Ioannidis, John Kastner, Anwar Mamat, et~al.
\newblock Cedar: A new language for expressive, fast, safe, and analyzable authorization.
\newblock {\em Proceedings of the ACM on Programming Languages}, 8(OOPSLA1):670--697, 2024.

\bibitem{debenedetti2025defeating}
Edoardo Debenedetti, Ilia Shumailov, Tianqi Fan, Jamie Hayes, Nicholas Carlini, Daniel Fabian, Christoph Kern, Chongyang Shi, Andreas Terzis, and Florian Tram{\`e}r.
\newblock Defeating prompt injections by design.
\newblock {\em arXiv preprint arXiv:2503.18813}, 2025.

\bibitem{debenedetti2024agentdojo}
Edoardo Debenedetti, Jie Zhang, Mislav Balunovi{\'c}, Luca Beurer-Kellner, Marc Fischer, and Florian Tram{\`e}r.
\newblock Agentdojo: A dynamic environment to evaluate attacks and defenses for llm agents.
\newblock {\em arXiv e-prints}, pages arXiv--2406, 2024.

\bibitem{aws-cognito}
Abrom Douglas.
\newblock Empower ai agents with user context using amazon cognito.
\newblock \url{https://aws.amazon.com/blogs/security/empower-ai-agents-with-user-context-using-amazon-cognito/}, n.d.

\bibitem{fawaz2026text}
Kassem Fawaz, Ren Yi, Octavian Suciu, Rishabh Khandelwal, Hamza Harkous, Nina Taft, and Marco Gruteser.
\newblock Text-based personas for simulating user privacy decisions.
\newblock {\em arXiv preprint arXiv:2603.19791}, 2026.

\bibitem{felt2011android}
Adrienne~Porter Felt, Erika Chin, Steve Hanna, Dawn Song, and David Wagner.
\newblock Android permissions demystified.
\newblock In {\em Proceedings of the 18th ACM conference on Computer and communications security}, pages 627--638, 2011.

\bibitem{fu2024imprompter}
Xiaohan Fu, Shuheng Li, Zihan Wang, Yihao Liu, Rajesh~K Gupta, Taylor Berg-Kirkpatrick, and Earlence Fernandes.
\newblock Imprompter: Tricking llm agents into improper tool use.
\newblock {\em arXiv preprint arXiv:2410.14923}, 2024.

\bibitem{ganter1999formal}
Bernhard Ganter and Rudolf Wille.
\newblock {\em Formal concept analysis}, volume 150.
\newblock Springer, 1999.

\bibitem{Gemini-workspace}
Google.
\newblock Connect google workspace apps \& services to gemini apps.
\newblock \url{https://support.google.com/gemini/answer/15229592}, n.d.

\bibitem{Gemini}
Google.
\newblock Gemini.
\newblock \url{https://gemini.google.com/}, n.d.

\bibitem{greshake2023not}
Kai Greshake, Sahar Abdelnabi, Shailesh Mishra, Christoph Endres, Thorsten Holz, and Mario Fritz.
\newblock Not what you've signed up for: Compromising real-world llm-integrated applications with indirect prompt injection.
\newblock In {\em Proceedings of the 16th ACM workshop on artificial intelligence and security}, pages 79--90, 2023.

\bibitem{han2024llm}
Shanshan Han, Qifan Zhang, Yuhang Yao, Weizhao Jin, and Zhaozhuo Xu.
\newblock Llm multi-agent systems: Challenges and open problems.
\newblock {\em arXiv preprint arXiv:2402.03578}, 2024.

\bibitem{rfc6749}
Dick Hardt.
\newblock {The OAuth 2.0 Authorization Framework}.
\newblock \url{https://www.rfc-editor.org/info/rfc6749}, October 2012.
\newblock URL: \url{https://www.rfc-editor.org/info/rfc6749}, \href {https://doi.org/10.17487/RFC6749} {\path{doi:10.17487/RFC6749}}.

\bibitem{he2025plan}
Gaole He, Gianluca Demartini, and Ujwal Gadiraju.
\newblock Plan-then-execute: An empirical study of user trust and team performance when using llm agents as a daily assistant.
\newblock In {\em Proceedings of the 2025 CHI Conference on Human Factors in Computing Systems}, pages 1--22, 2025.

\bibitem{hung2025attention}
Kuo-Han Hung, Ching-Yun Ko, Ambrish Rawat, I-Hsin Chung, Winston~H Hsu, and Pin-Yu Chen.
\newblock Attention tracker: Detecting prompt injection attacks in llms.
\newblock In {\em Findings of the Association for Computational Linguistics: NAACL 2025}, pages 2309--2322, 2025.

\bibitem{huq2025cowpilot}
Faria Huq, Zora~Zhiruo Wang, Frank~F Xu, Tianyue Ou, Shuyan Zhou, Jeffrey~P Bigham, and Graham Neubig.
\newblock Cowpilot: A framework for autonomous and human-agent collaborative web navigation.
\newblock {\em arXiv preprint arXiv:2501.16609}, 2025.

\bibitem{jha2025breaking}
Rishi Jha, Harold Triedman, Justin Wagle, and Vitaly Shmatikov.
\newblock Breaking and fixing defenses against control-flow hijacking in multi-agent systems.
\newblock {\em arXiv preprint arXiv:2510.17276}, 2025.

\bibitem{ji2026measuring}
Zimo Ji, Zongjie Li, Wenyuan Jiang, Yudong Gao, and Shuai Wang.
\newblock Measuring the permission gate: A stress-test evaluation of claude code's auto mode.
\newblock {\em arXiv preprint arXiv:2604.04978}, 2026.

\bibitem{kim2025prompt}
Juhee Kim, Woohyuk Choi, and Byoungyoung Lee.
\newblock Prompt flow integrity to prevent privilege escalation in llm agents.
\newblock {\em arXiv preprint arXiv:2503.15547}, 2025.

\bibitem{kirwan2020repetition}
C~Brock Kirwan, Daniel~K Bjornn, Bonnie~Brinton Anderson, Anthony Vance, David Eargle, and Jeffrey~L Jenkins.
\newblock Repetition of computer security warnings results in differential repetition suppression effects as revealed with functional mri.
\newblock {\em Frontiers in Psychology}, 11:528079, 2020.

\bibitem{kuntz2025harm}
Thomas Kuntz, Agatha Duzan, Hao Zhao, Francesco Croce, Zico Kolter, Nicolas Flammarion, and Maksym Andriushchenko.
\newblock Os-harm: A benchmark for measuring safety of computer use agents.
\newblock {\em arXiv preprint arXiv:2506.14866}, 2025.

\bibitem{langchain-bench}
LangChain.
\newblock Langchain benchmarks.
\newblock \url{https://langchain-ai.github.io/langchain-benchmarks/index.html\#}, n.d.

\bibitem{lau2000distributed}
Felix Lau, Stuart~H Rubin, Michael~H Smith, and Ljiljana Trajkovic.
\newblock Distributed denial of service attacks.
\newblock In {\em Smc 2000 conference proceedings. 2000 ieee international conference on systems, man and cybernetics.'cybernetics evolving to systems, humans, organizations, and their complex interactions'(cat. no. 0}, volume~3, pages 2275--2280. IEEE, 2000.

\bibitem{lee2023rlaif}
Harrison Lee, Samrat Phatale, Hassan Mansoor, Thomas Mesnard, Johan Ferret, Kellie Lu, Colton Bishop, Ethan Hall, Victor Carbune, Abhinav Rastogi, et~al.
\newblock Rlaif vs. rlhf: Scaling reinforcement learning from human feedback with ai feedback.
\newblock {\em arXiv preprint arXiv:2309.00267}, 2023.

\bibitem{levy2024tasktokensimpactinput}
Mosh Levy, Alon Jacoby, and Yoav Goldberg.
\newblock Same task, more tokens: the impact of input length on the reasoning performance of large language models.
\newblock In Lun-Wei Ku, Andre Martins, and Vivek Srikumar, editors, {\em Proceedings of the 62nd Annual Meeting of the Association for Computational Linguistics (Volume 1: Long Papers)}, pages 15339--15353, Bangkok, Thailand, August 2024. Association for Computational Linguistics.
\newblock URL: \url{https://aclanthology.org/2024.acl-long.818/}, \href {https://doi.org/10.18653/v1/2024.acl-long.818} {\path{doi:10.18653/v1/2024.acl-long.818}}.

\bibitem{li2025ace}
Evan Li, Tushin Mallick, Evan Rose, William Robertson, Alina Oprea, and Cristina Nita-Rotaru.
\newblock Ace: A security architecture for llm-integrated app systems.
\newblock {\em arXiv preprint arXiv:2504.20984}, 2025.

\bibitem{li2026agentdyn}
Hao Li, Ruoyao Wen, Shanghao Shi, Ning Zhang, and Chaowei Xiao.
\newblock Agentdyn: A dynamic open-ended benchmark for evaluating prompt injection attacks of real-world agent security system.
\newblock {\em arXiv preprint arXiv:2602.03117}, 2026.

\bibitem{li2007role}
Ninghui Li, Tiancheng Li, Ian Molloy, Qihua Wang, Elisa Bertino, Seraphin Calo, and Jorge Lobo.
\newblock Role mining for engineering and optimizing role based access control systems.
\newblock {\em Purdue University, IBM TJ Watson Research Center}, 2007.

\bibitem{liu2023prompt}
Yi~Liu, Gelei Deng, Yuekang Li, Kailong Wang, Zihao Wang, Xiaofeng Wang, Tianwei Zhang, Yepang Liu, Haoyu Wang, Yan Zheng, et~al.
\newblock Prompt injection attack against llm-integrated applications.
\newblock {\em arXiv preprint arXiv:2306.05499}, 2023.

\bibitem{liu2024formalizing}
Yupei Liu, Yuqi Jia, Runpeng Geng, Jinyuan Jia, and Neil~Zhenqiang Gong.
\newblock Formalizing and benchmarking prompt injection attacks and defenses.
\newblock In {\em 33rd USENIX Security Symposium (USENIX Security 24)}, pages 1831--1847, 2024.

\bibitem{lu2025toolsandbox}
Jiarui Lu, Thomas Holleis, Yizhe Zhang, Bernhard Aumayer, Feng Nan, Haoping Bai, Shuang Ma, Shen Ma, Mengyu Li, Guoli Yin, et~al.
\newblock Toolsandbox: A stateful, conversational, interactive evaluation benchmark for llm tool use capabilities.
\newblock In {\em Findings of the Association for Computational Linguistics: NAACL 2025}, pages 1160--1183, 2025.

\bibitem{luo2025code}
Weidi Luo, Qiming Zhang, Tianyu Lu, Xiaogeng Liu, Bin Hu, Hung-Chun Chiu, Siyuan Ma, Yizhe Zhang, Xusheng Xiao, Yinzhi Cao, et~al.
\newblock Code agent can be an end-to-end system hacker: Benchmarking real-world threats of computer-use agent.
\newblock {\em arXiv preprint arXiv:2510.06607}, 2025.

\bibitem{malkin2022runtime}
Nathan Malkin, David Wagner, and Serge Egelman.
\newblock Runtime permissions for privacy in proactive intelligent assistants.
\newblock In {\em Eighteenth Symposium on Usable Privacy and Security (SOUPS 2022)}, pages 633--651, 2022.

\bibitem{pixel-ai}
Molly McHugh-Johnson.
\newblock 14 new things you can do with pixel thanks to ai.
\newblock \url{https://blog.google/products/pixel/google-pixel-9-new-ai-features/}, n.d.

\bibitem{ms-mcp-risk}
Microsoft.
\newblock Plug, play, and prey: The security risks of the model context protocol.
\newblock \url{https://techcommunity.microsoft.com/blog/microsoftdefendercloudblog/plug-play-and-prey-the-security-risks-of-the-model-context-protocol/4410829}, n.d.

\bibitem{ms-wassette}
Microsoft.
\newblock Wassette: A security-oriented runtime that runs webassembly components via mcp.
\newblock \url{https://microsoft.github.io/wassette/latest/}, n.d.

\bibitem{mozannar2025magentic}
Hussein Mozannar, Gagan Bansal, Cheng Tan, Adam Fourney, Victor Dibia, Jingya Chen, Jack Gerrits, Tyler Payne, Matheus~Kunzler Maldaner, Madeleine Grunde-McLaughlin, et~al.
\newblock Magentic-ui: Towards human-in-the-loop agentic systems.
\newblock {\em arXiv preprint arXiv:2507.22358}, 2025.

\bibitem{magentic-ui}
Hussein Mozannar, Gagan Bansal, Cheng Tan, Adam Fourney, Victor Dibia, Jingya Chen, Jack Gerrits, Tyler Payne, Matheus~Kunzler Maldaner, Madeleine Grunde-McLaughlin, Erkang~(Eric) Zhu, Griffin Bassman, Jacob Alber, Peter Chang, Ricky Loynd, Friederike Niedtner, Ece Kamar, Maya Murad, Rafah Hosn, and Saleema Amershi.
\newblock Magentic-ui, an experimental human-centered web agent.
\newblock \url{https://www.microsoft.com/en-us/research/blog/magentic-ui-an-experimental-human-centered-web-agent}, n.d.

\bibitem{codex-auto-mode}
Phemex News.
\newblock Openai launches auto-review mode for codex.
\newblock \url{https://phemex.com/news/article/openai-launches-autoreview-mode-for-codex-75603}, n.d.

\bibitem{ChatGPT}
OpenAI.
\newblock Chatgpt.
\newblock \url{https://chat.openai.com/}, n.d.

\bibitem{chagpt-cua}
OpenAI.
\newblock Computer use agent.
\newblock \url{https://platform.openai.com/docs/guides/tools-computer-use}, n.d.

\bibitem{ChatGPT-conn}
OpenAI.
\newblock Connectors and actions terms.
\newblock \url{https://openai.com/policies/connectors-actions-terms/}, n.d.

\bibitem{chagpt-operator}
OpenAI.
\newblock Introducing operator.
\newblock \url{https://openai.com/index/introducing-operator/}, n.d.

\bibitem{openrouter}
OpenRouter.
\newblock The unified interface for llms.
\newblock \url{https://openrouter.ai/}, n.d.

\bibitem{ouyang2022training}
Long Ouyang, Jeffrey Wu, Xu~Jiang, Diogo Almeida, Carroll Wainwright, Pamela Mishkin, Chong Zhang, Sandhini Agarwal, Katarina Slama, Alex Ray, et~al.
\newblock Training language models to follow instructions with human feedback.
\newblock {\em Advances in neural information processing systems}, 35:27730--27744, 2022.

\bibitem{owasp-excessive}
OWASP.
\newblock Llm06:2025 excessive agency.
\newblock \url{https://genai.owasp.org/llmrisk/llm062025-excessive-agency/}.

\bibitem{owasp-privilege-escalation}
OWASP.
\newblock Testing for privilege escalation.
\newblock \url{https://owasp.org/www-project-web-security-testing-guide/stable/4-Web_Application_Security_Testing/05-Authorization_Testing/03-Testing_for_Privilege_Escalation}, n.d.

\bibitem{piet2024jatmo}
Julien Piet, Maha Alrashed, Chawin Sitawarin, Sizhe Chen, Zeming Wei, Elizabeth Sun, Basel Alomair, and David Wagner.
\newblock Jatmo: Prompt injection defense by task-specific finetuning.
\newblock In {\em European Symposium on Research in Computer Security}, pages 105--124. Springer, 2024.

\bibitem{prabhakar2025apigen}
Akshara Prabhakar, Zuxin Liu, Ming Zhu, Jianguo Zhang, Tulika Awalgaonkar, Shiyu Wang, Zhiwei Liu, Haolin Chen, Thai Hoang, Juan~Carlos Niebles, et~al.
\newblock Apigen-mt: Agentic pipeline for multi-turn data generation via simulated agent-human interplay.
\newblock {\em arXiv preprint arXiv:2504.03601}, 2025.

\bibitem{rawte2023survey}
Vipula Rawte, Amit Sheth, and Amitava Das.
\newblock A survey of hallucination in large foundation models.
\newblock {\em arXiv preprint arXiv:2309.05922}, 2023.

\bibitem{sandhu1998role}
Ravi~S Sandhu.
\newblock Role-based access control.
\newblock In {\em Advances in computers}, volume~46, pages 237--286. Elsevier, 1998.

\bibitem{shavit2023practices}
Yonadav Shavit, Sandhini Agarwal, Miles Brundage, Steven Adler, Cullen O’Keefe, Rosie Campbell, Teddy Lee, Pamela Mishkin, Tyna Eloundou, Alan Hickey, et~al.
\newblock Practices for governing agentic ai systems.
\newblock {\em Research Paper, OpenAI}, 2023.

\bibitem{shen2021can}
Bingyu Shen, Lili Wei, Chengcheng Xiang, Yudong Wu, Mingyao Shen, Yuanyuan Zhou, and Xinxin Jin.
\newblock Can systems explain permissions better? understanding users' misperceptions under smartphone runtime permission model.
\newblock In {\em 30th USENIX Security Symposium (USENIX Security 21)}, pages 751--768, 2021.

\bibitem{shi2025prompt}
Jiawen Shi, Zenghui Yuan, Guiyao Tie, Pan Zhou, Neil~Zhenqiang Gong, and Lichao Sun.
\newblock Prompt injection attack to tool selection in llm agents.
\newblock {\em arXiv preprint arXiv:2504.19793}, 2025.

\bibitem{shi2025progent}
Tianneng Shi, Jingxuan He, Zhun Wang, Linyu Wu, Hongwei Li, Wenbo Guo, and Dawn Song.
\newblock Progent: Programmable privilege control for llm agents.
\newblock {\em arXiv preprint arXiv:2504.11703}, 2025.

\bibitem{singhal2023large}
Karan Singhal, Shekoofeh Azizi, Tao Tu, S~Sara Mahdavi, Jason Wei, Hyung~Won Chung, Nathan Scales, Ajay Tanwani, Heather Cole-Lewis, Stephen Pfohl, et~al.
\newblock Large language models encode clinical knowledge.
\newblock {\em Nature}, 620(7972):172--180, 2023.

\bibitem{south2025authenticated}
Tobin South, Samuele Marro, Thomas Hardjono, Robert Mahari, Cedric~Deslandes Whitney, Dazza Greenwood, Alan Chan, and Alex Pentland.
\newblock Authenticated delegation and authorized ai agents.
\newblock {\em arXiv preprint arXiv:2501.09674}, 2025.

\bibitem{openclaw}
Peter Steinberger.
\newblock Openclaw: Personal ai assistant.
\newblock \url{https://github.com/openclaw/openclaw}, n.d.

\bibitem{google-a2a}
Rao Surapaneni, Miku Jha, Michael Vakoc, and Todd Segal.
\newblock Announcing the agent2agent protocol (a2a).
\newblock \url{https://developers.googleblog.com/en/a2a-a-new-era-of-agent-interoperability/}, 2025.

\bibitem{syros2025saga}
Georgios Syros, Anshuman Suri, Cristina Nita-Rotaru, and Alina Oprea.
\newblock Saga: A security architecture for governing ai agentic systems.
\newblock {\em arXiv preprint arXiv:2504.21034}, 2025.

\bibitem{toyer2023tensor}
Sam Toyer, Olivia Watkins, Ethan~Adrian Mendes, Justin Svegliato, Luke Bailey, Tiffany Wang, Isaac Ong, Karim Elmaaroufi, Pieter Abbeel, Trevor Darrell, et~al.
\newblock Tensor trust: Interpretable prompt injection attacks from an online game.
\newblock {\em arXiv preprint arXiv:2311.01011}, 2023.

\bibitem{triedman2025multi}
Harold Triedman, Rishi Jha, and Vitaly Shmatikov.
\newblock Multi-agent systems execute arbitrary malicious code.
\newblock {\em arXiv preprint arXiv:2503.12188}, 2025.

\bibitem{trivedi2024appworld}
Harsh Trivedi, Tushar Khot, Mareike Hartmann, Ruskin Manku, Vinty Dong, Edward Li, Shashank Gupta, Ashish Sabharwal, and Niranjan Balasubramanian.
\newblock Appworld: A controllable world of apps and people for benchmarking interactive coding agents.
\newblock In {\em Proceedings of the 62nd Annual Meeting of the Association for Computational Linguistics (Volume 1: Long Papers)}, pages 16022--16076, 2024.

\bibitem{tsai2025contextual}
Lillian Tsai and Eugene Bagdasarian.
\newblock Contextual agent security: A policy for every purpose.
\newblock In {\em Proceedings of the 2025 Workshop on Hot Topics in Operating Systems}, pages 8--17, 2025.

\bibitem{pocketos-delete}
Mark Tyson.
\newblock Claude-powered ai coding agent deletes entire company database in 9 seconds.
\newblock \url{https://www.tomshardware.com/tech-industry/artificial-intelligence/claude-powered-ai-coding-agent-deletes-entire-company-database-in-9-seconds-backups-zapped-after-cursor-tool-powered-by-anthropics-claude-goes-rogue}, n.d.

\bibitem{browser-use}
Browser Use.
\newblock Enable ai to control your browser.
\newblock \url{https://github.com/browser-use/browser-use}, n.d.

\bibitem{wallace2024instruction}
Eric Wallace, Kai Xiao, Reimar Leike, Lilian Weng, Johannes Heidecke, and Alex Beutel.
\newblock The instruction hierarchy: Training llms to prioritize privileged instructions.
\newblock {\em arXiv preprint arXiv:2404.13208}, 2024.

\bibitem{wang2025agentspec}
Haoyu Wang, Christopher~M Poskitt, and Jun Sun.
\newblock Agentspec: Customizable runtime enforcement for safe and reliable llm agents.
\newblock {\em arXiv preprint arXiv:2503.18666}, 2025.

\bibitem{wang2023plan}
Lei Wang, Wanyu Xu, Yihuai Lan, Zhiqiang Hu, Yunshi Lan, Roy Ka-Wei Lee, and Ee-Peng Lim.
\newblock Plan-and-solve prompting: Improving zero-shot chain-of-thought reasoning by large language models.
\newblock In {\em Proceedings of the 61st annual meeting of the association for computational linguistics (volume 1: long papers)}, pages 2609--2634, 2023.

\bibitem{wei2023jailbroken}
Alexander Wei, Nika Haghtalab, and Jacob Steinhardt.
\newblock Jailbroken: How does llm safety training fail?
\newblock {\em Advances in Neural Information Processing Systems}, 36:80079--80110, 2023.

\bibitem{wijesekera2017feasibility}
Primal Wijesekera, Arjun Baokar, Lynn Tsai, Joel Reardon, Serge Egelman, David Wagner, and Konstantin Beznosov.
\newblock The feasibility of dynamically granted permissions: Aligning mobile privacy with user preferences.
\newblock In {\em 2017 IEEE Symposium on Security and Privacy (SP)}, pages 1077--1093. IEEE, 2017.

\bibitem{dual-llm}
Simon Willison.
\newblock The dual llm pattern for building ai assistants that can resist prompt injection.
\newblock \url{https://simonwillison.net/2023/Apr/25/dual-llm-pattern/}, n.d.

\bibitem{wu2024system}
Fangzhou Wu, Ethan Cecchetti, and Chaowei Xiao.
\newblock System-level defense against indirect prompt injection attacks: An information flow control perspective.
\newblock {\em arXiv preprint arXiv:2409.19091}, 2024.

\bibitem{wu2025isolategpt}
Yuhao Wu, Franziska Roesner, Tadayoshi Kohno, Ning Zhang, and Umar Iqbal.
\newblock Isolategpt: An execution isolation architecture for llm-based agentic systems.
\newblock In {\em NDSS}, 2025.

\bibitem{wu2025towards}
Yuhao Wu, Ke~Yang, Franziska Roesner, Tadayoshi Kohno, Ning Zhang, and Umar Iqbal.
\newblock Towards automating data access permissions in ai agents.
\newblock {\em arXiv preprint arXiv:2511.17959}, 2025.

\bibitem{xu2024safedecoding}
Zhangchen Xu, Fengqing Jiang, Luyao Niu, Jinyuan Jia, Bill~Yuchen Lin, and Radha Poovendran.
\newblock Safedecoding: Defending against jailbreak attacks via safety-aware decoding.
\newblock {\em arXiv preprint arXiv:2402.08983}, 2024.

\bibitem{xu2026storage}
Zhenlin Xu, Xiaogang Zhu, Yu~Yao, Minhui Xue, and Yiliao Song.
\newblock From storage to steering: Memory control flow attacks on llm agents.
\newblock {\em arXiv preprint arXiv:2603.15125}, 2026.

\bibitem{xu2026unlocking}
Zhihao Xu, Rumei Li, Jiahuan Li, Rongxiang Weng, Jingang Wang, Xunliang Cai, and Xiting Wang.
\newblock Unlocking implicit experience: Synthesizing tool-use trajectories from text.
\newblock {\em arXiv preprint arXiv:2601.10355}, 2026.

\bibitem{yao2024tau}
Shunyu Yao, Noah Shinn, Pedram Razavi, and Karthik Narasimhan.
\newblock $\tau$-bench: A benchmark for tool-agent-user interaction in real-world domains.
\newblock {\em arXiv preprint arXiv:2406.12045}, 2024.

\bibitem{yao2022react}
Shunyu Yao, Jeffrey Zhao, Dian Yu, Nan Du, Izhak Shafran, Karthik~R Narasimhan, and Yuan Cao.
\newblock React: Synergizing reasoning and acting in language models.
\newblock In {\em The eleventh international conference on learning representations}, 2022.

\bibitem{yi2025privacy}
Ren Yi, Octavian Suciu, Adria Gascon, Sarah Meiklejohn, Eugene Bagdasarian, and Marco Gruteser.
\newblock Privacy reasoning in ambiguous contexts.
\newblock {\em arXiv preprint arXiv:2506.12241}, 2025.

\bibitem{zhang2024agent}
Hanrong Zhang, Jingyuan Huang, Kai Mei, Yifei Yao, Zhenting Wang, Chenlu Zhan, Hongwei Wang, and Yongfeng Zhang.
\newblock Agent security bench (asb): Formalizing and benchmarking attacks and defenses in llm-based agents.
\newblock {\em arXiv preprint arXiv:2410.02644}, 2024.

\bibitem{zhang2411attacking}
Yanzhe Zhang, Tao Yu, and Diyi Yang.
\newblock Attacking vision-language computer agents via pop-ups.
\newblock {\em arXiv preprint arXiv:2411.02391}, 2024.

\bibitem{zhang2025siren}
Yue Zhang, Yafu Li, Leyang Cui, Deng Cai, Lemao Liu, Tingchen Fu, Xinting Huang, Enbo Zhao, Yu~Zhang, Yulong Chen, et~al.
\newblock Siren’s song in the ai ocean: A survey on hallucination in large language models.
\newblock {\em Computational Linguistics}, pages 1--45, 2025.

\bibitem{zou2023universal}
Andy Zou, Zifan Wang, Nicholas Carlini, Milad Nasr, J~Zico Kolter, and Matt Fredrikson.
\newblock Universal and transferable adversarial attacks on aligned language models.
\newblock {\em arXiv preprint arXiv:2307.15043}, 2023.

\bibitem{zou2025poisonedrag}
Wei Zou, Runpeng Geng, Binghui Wang, and Jinyuan Jia.
\newblock $\{$PoisonedRAG$\}$: Knowledge corruption attacks to $\{$Retrieval-Augmented$\}$ generation of large language models.
\newblock In {\em 34th USENIX Security Symposium (USENIX Security 25)}, pages 3827--3844, 2025.

\bibitem{zverev2025aside}
Egor Zverev, Evgenii Kortukov, Alexander Panfilov, Alexandra Volkova, Soroush Tabesh, Sebastian Lapuschkin, Wojciech Samek, and Christoph~H Lampert.
\newblock Aside: Architectural separation of instructions and data in language models.
\newblock {\em arXiv preprint arXiv:2503.10566}, 2025.

\end{thebibliography}

\cleardoublepage

\section{Algorithm}

\begin{algorithm}[]
\caption{BUILD-TREE($S, \mathrm{Methods}$)}
\label{alg:build-scope-tree}
\KwIn{
    Scopes $S$; function $\mathrm{Methods}(s)$ returning allowed methods of $s$
}
\KwOut{
    Directed graph $\mathbb{T} = (V, E)$
}

\ForEach{$s \in S$}{
    $M_s \gets \mathrm{Methods}(s)$\;
    $\mathit{Candidates} \gets \varnothing$\;

    \ForEach{$s' \in S \setminus \{s\}$}{
        $M_{s'} \gets \mathrm{Methods}(s')$\;
        \If{$M_s \subset M_{s'}$}{        %
            $\mathit{Candidates} \gets \mathit{Candidates} \cup \{s'\}$\;
        }
    }

   \If{$\mathit{Candidates} = \varnothing$}{
    $\mathit{Parents}[s] \gets \varnothing$\;
    }
    \Else{
        $\mathit{Parents}[s] \gets
            \arg\min_{s' \in \mathit{Candidates}} |\mathrm{Methods}(s')|$\;
}

}

$V \gets S$\;
$E \gets \varnothing$\;

\ForEach{$s \in S$}{
    \ForEach{$p \in \mathit{Parents}[s]$}{
        $E \gets E \cup \{(p, s)\}$\;
    }
}

\Return $(V, E)$\;

\end{algorithm}

This algorithm (Alg.~\ref{alg:build-scope-tree}) constructs a permission hierarchy from a set of scopes $S$ and a function $\mathrm{Methods}(s)$, which returns the set of API methods permitted by each scope $s$.
The goal is to produce a directed graph $\mathbb{T} = (V, E)$ in which each edge $(p, s)$ indicates that $p$ is a strictly more permissive scope than $s$,
and that $p$ is a \emph{minimal} such superset in terms of allowed methods.
Consider an example from the Google Calendar API.
Let $s_1 = \texttt{calendar.events}$ and
$s_2 = \texttt{calendar.events.owned}$. The scope $s_1$ permits a certain set
of event-level operations, while $s_2$ permits all operations of $s_1$ plus
additional owner-level actions.
In this case,
\[
    \mathrm{Methods}(s_1) \subset \mathrm{Methods}(s_2),
\]
so $s_2$ is a candidate parent of $s_1$. In the constructed graph, this
relationship appears as a directed edge
\[
    (s_2,\, s_1) \in E ,
\]
representing that $s_2$ is strictly more permissive than $s_1$.

The algorithm determines the parent set for each scope $s \in S$ by comparing
$\mathrm{Methods}(s)$ with $\mathrm{Methods}(s')$ for all $s' \in S \setminus
\{s\}$. Any scope $s'$ satisfying
$\mathrm{Methods}(s) \subset \mathrm{Methods}(s')$ is collected into the set of
candidate parents. If this set is empty, $s$ has no parent. Otherwise, the
algorithm selects the ``closest'' superscopes by choosing those candidates whose
method sets have minimum size. Formally,
\[
    \mathit{Parents}[s] \gets
    \arg\min_{s' \in \mathit{Candidates}}
        \bigl|\mathrm{Methods}(s')\bigr| \, ,
\]
ensuring that each parent differs from $s$ by the smallest possible increase in
permitted functionality.

Finally, the graph $\mathbb{T} = (V, E)$ is assembled by taking $V = S$ and adding an edge $(p, s)$ for every $p \in \mathit{Parents}[s]$. 
The resulting structure organizes scopes (permissions) according to their relative sensitivity, forming a clean hierarchy suitable for downstream least-privilege analysis.

{
\begin{table*}[!t]
\centering
\begin{tabular}{l|ccc|ccc} 
\toprule
\textbf{Applications}  & \multicolumn{3}{c|}{\textbf{Single-method}} & \multicolumn{3}{c}{\textbf{Multi-method}} \\
& \# Requests & $C_{scope}$ & $C_{method}$ & \# Requests & $C_{scope}$ & $C_{method}$ \\
\midrule
Gmail  & 79 & 100\% & 100\% & 151 &100\% &100\% \\
Google Calendar   & 37 & 62.5\% & 100\% & 117 & 68.8\% &100\% \\
Google Drive   & 58 & 100\% & 100\% & 168 &100\% &98.2\% \\
Google Storage  & 81 &  100\%  & 100\% & 114 & 100\%&100\% \\
Slack  & 247 &  100\%  & 100\% & 190 &100\% & 91.4\% \\
Dropbox  & 120 &  100\%  & 100\% & 194 &100\% & 100\%\\
Outlook   & 44 & 100\% & 84.1\%  & 154 &79.5\% &100\% \\
Outlook Calendar  & 51 &  100\% & 82.4\% & 113 &85.7\% &82.4\% \\
Notion  & 32 &  100\%  & 100\% & 127 &100\% & 100\%\\
Zoom  & 184 & 100\% & 100\% & 161 & 95.7\% & 97.8\% \\
\midrule
Suite 1  & - & - & - & 77  & 94.7\% & 83.3\%\\
Suite 2  & - & - & - & 93 & 33.3\% & 15.1\% \\
\bottomrule
\end{tabular}
\caption{Statistics for our synthetic dataset. For each application, we report the number of generated requests together with their coverage on scopes ($C_{Scope}$) and methods ($C_{Method}$) for both single-method and multi-method scenarios.}
\label{tab:api-scopes-coverage}
\end{table*}
}

\begin{figure*}[!t]
    \centering
    \includegraphics[width=1.\linewidth]{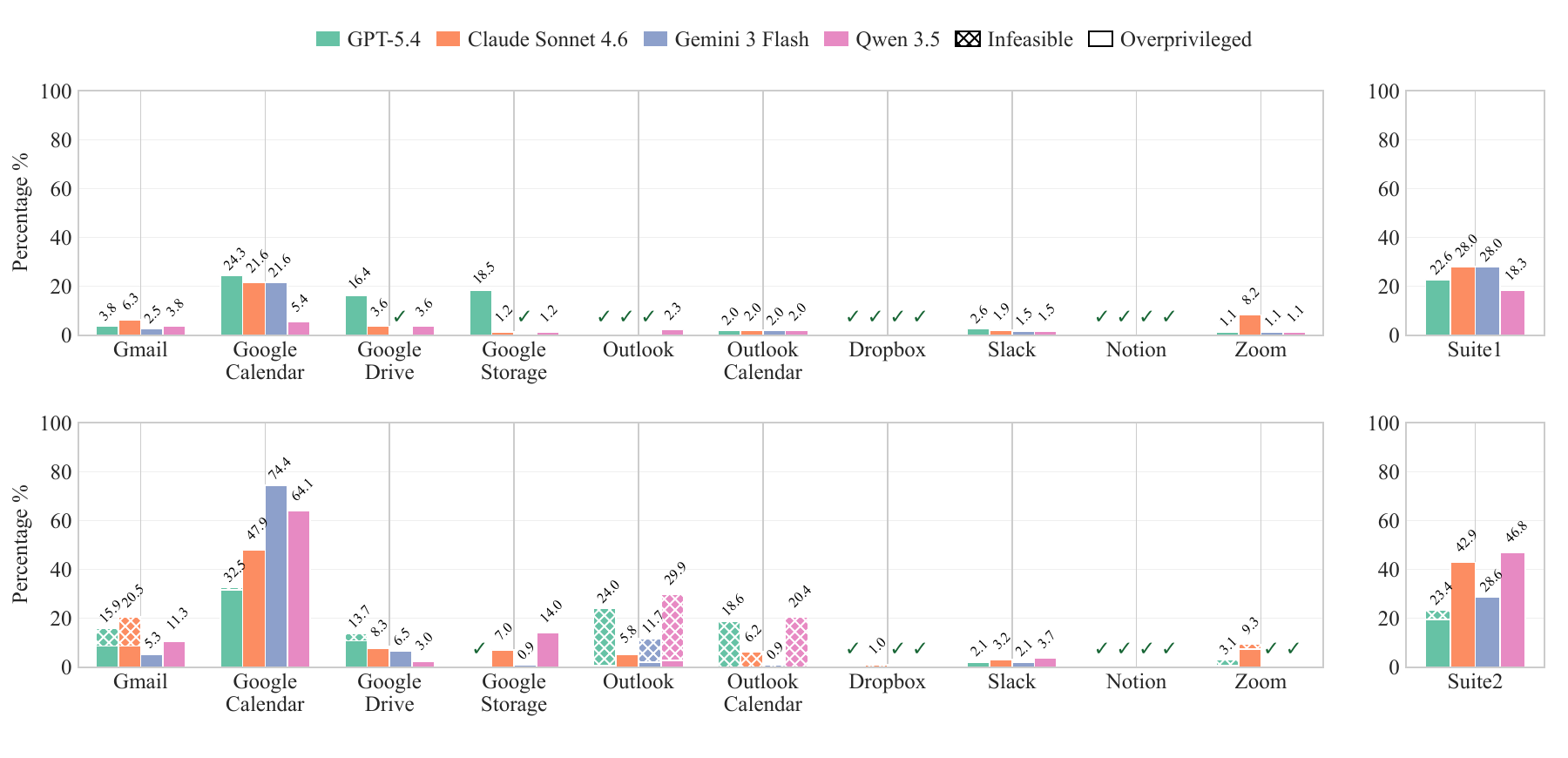}
    \caption{Percentage of permission mismatches for the LLM-based permission inference baseline relative to \sys. Top left: single-application, single-method. Bottom left: single-application, multi-method. Right: multi-application, multi-method. Green checkmarks indicate zero mismatches.}
    \label{fig:misalignment-detail}
\end{figure*} 

\section{Evaluation Details}

\subsection{Reconstructed Permission Hierarchies}
\label{appendix:hierarchies}

In the reconstructed permission hierarchies, we find that applications exhibit very different structural patterns. 
We present these differences for each application by reporting the maximum and average heights of the reconstructed hierarchies in \tab{tab:api-scopes}. 
Some applications, such as Gmail, Google Drive, and Google Calendar, form deep hierarchical graphs that condense into proper trees once cycles are removed. 
For example, Google Calendar forms a single tree of height~5. 
In contrast, applications such as Slack, Zoom, and Dropbox have relatively shallow and fragmented hierarchies composed of many disconnected components.
Many components are either small trees or single scopes, forming shallow forests with average heights near 1.0.
These contrasting structures reflect different design philosophies. 
Some services favor simplicity and define a small number of broad permission groups, including highly privileged scopes, which naturally appear as roots in the reconstructed hierarchy. 
Others pursue fine-grained access control, avoiding broad scopes and instead exposing many low-privilege granular scopes, which leads to flat and fragmented hierarchies.

\subsection{Coverage of Requests}
\label{appendix:permissioner}
\tab{tab:api-scopes-coverage} summarizes the statistics for each application and the corresponding generated requests. The generated requests achieve high coverage across all scenarios except for Suite 2, which we attribute to the complexity of the applications it contains.

\subsection{Breakdown of LLM-inferred Permissions}
\label{appendix:permissioner-breakdown}
\fig{fig:misalignment-detail} provides a detailed breakdown of the percentage of infeasible and overprivileged permissions for the LLM-based permission inference baseline relative to \sys across all three scenarios.
For applications such as Google Calendar, Google Drive, and Gmail, the permission hierarchies are deep and tree-structured, and even in the single-application, single-method setting, the mismatch rate can exceed 24\%. 
In contrast, for applications with flat hierarchies, where each method maps to exactly one permission group, identifying the minimal set of permissions is straightforward, and the mismatch rate is very low (even zero), as seen in Dropbox, Outlook and Zoom.
Strict overprivilege further increases when moving from single-method to multi-method tasks, because the LLM must compose permissions across several required methods.
For Google Calendar, this effect leads to up to 74.4\% strict overprivilege in the multi-method setting.

\end{document}